\documentclass{aa}  

\usepackage{natbib}
\usepackage{graphicx}
\usepackage{subfigure}
\usepackage[varg]{txfonts}
\usepackage[colorlinks=true,linkcolor=blue,citecolor=blue,filecolor=blue,urlcolor=blue]{hyperref}  
\usepackage{CJKutf8}

\usepackage{hyperref}

\usepackage{graphicx}	
\usepackage{amsmath}	
\usepackage{amssymb}	
\usepackage[version=4]{mhchem}

\makeatletter
\def\uwave{\bgroup \markoverwith{\lower3.5\p@\hbox{\sixly \textcolor{red}{\char58}}}\ULon}
\font\sixly=lasy6 
\makeatother

\usepackage{bm}
\usepackage{color}
\usepackage{ulem}
\usepackage{xspace}

\definecolor{mygray}{gray}{0.6}
\definecolor{orange}{rgb}{1.0, 0.4, 0.0}
\definecolor{myblue}{rgb}{0.1, 0.5, 0.7}
\definecolor{LimeGreen}{rgb}{0.55, 0.78, 0.24}

\definecolor{mygray}{gray}{0.6}
\definecolor{orange}{rgb}{1.0, 0.4, 0.0}
\definecolor{purple}{rgb}{0.75,0.25,0.75}

\newcommand{\xxx}[1]{\textcolor{orange}{\textbf{xxx}\xspace}}

\newcommand{\ppp}[3]{$P_\mathrm{#1}{-}P_\mathrm{#2}{-}P_\mathrm{#3}$\xspace}

\newcommand{\fg}[1]{\mbox{Figure~\ref{fig:#1}}}
\newcommand{\Fg}[1]{\mbox{Figure~\ref{fig:#1}}}

\newcommand{\eq}[1]{Equation~(\ref{eq:#1})\xspace}
\newcommand{\Eq}[1]{Equation~(\ref{eq:#1})\xspace}

\newcommand{\tb}[1]{Table~\ref{tab:#1}\xspace}
\newcommand{\se}[1]{Section~\ref{sec:#1}\xspace}
\newcommand{\Se}[1]{Section~\ref{sec:#1}\xspace}
\newcommand{\ses}[2]{Sects.\ \ref{sec:#1} and \ref{sec:#2}}

\newcommand{\App}[1]{Appendix~\ref{app:#1}\xspace}

\newcommand{\bce}[1]{(b, c, e)}
\newcommand{\bcd}[1]{(b, c, d)}
\newcommand{\cde}[1]{(c, d, e)}

\usepackage{ulem}
\makeatletter
\def\uwave{\bgroup \markoverwith{\lower3.5\p@\hbox{\sixly \textcolor{red}{\char58}}}\ULon}
\font\sixly=lasy6 
\makeatother

\begin{document}


\title{The Dynamical History of the Kepler-221 Planet System}
\authorrunning{Yi, Ormel, Huang, Petit}
\author{Tian Yi (\begin{CJK*}{UTF8}{gbsn}易天\end{CJK*})
	\inst{1}
	\and
	Chris W. Ormel\inst{1}
	\and
	Shuo Huang (\begin{CJK*}{UTF8}{gbsn}黄硕\end{CJK*}) \inst{1,2}
	\and
	Antoine C. Petit\inst{3}
}
\institute{
	Department of Astronomy, Tsinghua University, Haidian DS 100084, Beijing, China \\
	\email{chrisormel@tsinghua.edu.cn}
	\and
	Leiden Observatory, Leiden University, P.O. Box 9513, 2300 RA Leiden, The Netherlands
	\and
	Universit\'e Cote d’Azur, CNRS, Observatoire de la Cote d’Azur, Laboratoire Lagrange, Nice, France
}

\date{Accepted XXX. Received YYY; in original form ZZZ}

\abstract{Kepler-221 is a G-type star hosting four planets. In this system, planets b, c, and e are in (or near) a 6:3:1 three-body resonance even though the planets' period ratios show significant departures from exact two-body commensurability. Importantly, the intermediate planet d is not part of the resonance chain. To reach this resonance configuration, we propose a scenario in which there were originally five planets in the system in a chain of first-order resonances. After disk dispersal, the resonance chain became unstable, and two planets quickly merged to become the current planet d. In addition, the \bce{} three-body resonance was re-established. We ran N-body simulations using \texttt{REBOUND} to investigate the parameter space under which this scenario can operate. We find that our envisioned scenario is possible when certain conditions are met. First, the reformation of the three-body resonance after planet merging requires convergent migration between planets b and c. Second, as has been previously pointed out, an efficient damping mechanism must operate to power the expansion of the \bce{} system. We find that planet d plays a crucial role during the orbital expansion phase due to destabilizing encounters of a three-body resonance between c, d, and e. A successful orbital expansion phase puts constraints on the planet properties in the Kepler-221 system including the planet mass ratios and the tidal quality factors for the planets. Our model can also be applied to other planet systems in resonance, such as Kepler-402 and K2-138. }


\keywords{Stars: individual: Kepler-221 - Planetary systems - Planets and satellites: dynamical evolution and stability - Planets and satellites: formation }
\maketitle


\section{Introduction}

In the past two decades, with various transit and radial-velocity surveys, the number of multi-exoplanetary systems detected has exceeded 900.\footnote{https://exoplanetarchive.ipac.caltech.edu} Among these, there is an excess of systems with adjacent planets in period ratios close to integer \citep{FabryckyEtal2014, SteffenHwang2015, HuangOrmel2023, HamerSchlaufman2024, DaiEtal2024}, which are often identified with mean-motion resonances (MMRs). The MMRs in these multi-planetary systems are believed to form early, in the gas-rich disk phase, through Type-I convergent migration \citep{GoldreichTremaine1979, LinPapaloizou1979}, with the exact resonant state related to the disk and planet properties \citep{OgiharaKobayashi2013, KajtaziEtal2023, HuangOrmel2023, BatyginPetit2023i}. Therefore, resonant systems preserve relics of formation processes and can unveil the footprint of the dynamical history of multi-planetary systems \citep{SnellgroveEtal2001, PapaloizouSzuszkiewicz2005, TeyssandierLibert2020, HuangOrmel2022, PichierriEtal2024i}.

The formal dynamical identification of a two-body resonance (2BR) is in terms of an angle \citep{MurrayEtal1999}, which relies on a precise knowledge of the argument of pericenter and is observationally hard to assess. However, it is much easier to recognize three planets in a three-body resonance (3BR) because the corresponding resonance angle only depends on the angular positions of the planets (their mean longitudes). Multiple exoplanetary systems have been confirmed to host planets in 3BRs, including GJ 876 \citep{RiveraEtal2010}, Kepler-80 \citep{Xie2013, MacDonaldEtal2016, MacDonaldEtal2021}, Kepler-221 \citep{RoweEtal2014, GoldbergBatygin2021}, Kepler-60 \citep{GozdziewskiEtal2016}, Kepler-223 \citep{MillsEtal2016}, TRAPPIST-1 \citep{GillonEtal2017, LugarEtal2017, AgolEtal2021, HuangOrmel2022}, K2-138 \citep{ChristiansenEtal2018, LeleuEtal2019, LeleuEtal2021, CerioniBeauge2023}, HD-158259 \citep{HaraEtal2020}, TOI-178 \citep{LeleuEtal2021TOI}, TOI-1136 \citep{DaiEtal2023}, HD-110067 \citep{LuqueEtal2023, LammersWinn2024}, and Kepler-402 (see \se{other_system}). 

\begin{table*}
\caption{\label{tab:Kepler-221}Dynamical configuration of Kepler-221 planetary system according to \citet{RoweEtal2014, BergerEtal2018}. }
\centering
\small
\renewcommand{\arraystretch}{1.2}
\begin{tabular}{ccccc}
\hline
Planet                       & b & c & d & e \\ \hline
Radius($R_\oplus$)                     &  
	$1.587^{+0.468}_{-0.121}$ & $2.963^{+0.467}_{-0.174}$   &  $2.835^{+0.425}_{-0.321}$   &  $3.415^{+0.151}_{-0.787}$  \\
Period(days)                   & 
	2.796  & 5.691 &  
	10.042 &  
	18.370 \\
Period Ratio to Inner Planet & \textbackslash{}  & 2.035  & 1.765  & 1.829  \\ 
$(p,q,r)$ of 3BR\tablefootmark{a}    &   &   (5,12,17) &   (3,5,8) \\
Normalized B-values of 3BR\tablefootmark{b} & & $1.90\times10^{-2}$ & $2.44\times10^{-2}$ \\ \hline
\end{tabular}
\tablefoot{ 
\tablefoottext{a}{Specify the closest $(p,q,r)$ 3BR (3BR with the smallest normalized B-value defined in \eq{B_norm}) among adjacent planets (i-1,i,i+1).} \\
\tablefoottext{b}{The non-adjacent planets b, c, and e are close to $(2,3,5)$ 3BR with a small normalized B-value of $1.03\times10^{-5}$.}
}
\vspace{0.1cm}

\end{table*}

Among the planetary systems with a 3BR, Kepler-221 is unique because it hosts a non-adjacent 3BR containing a second-order 2BR \citep{GoldbergBatygin2021}. Kepler-221 is a G5-type star ($R_s=0.82R_\odot$) hosting four planets (b, c, d, and e) at a distance of 385 $\mathrm{pc}$, with the system information listed in \tb{Kepler-221} \citep{RoweEtal2014, BergerEtal2018}. Kepler-221 system is likely to be relatively young, with its large lithium abundance suggesting that the system is younger than the Hyades with an age estimated around 650 Myr \citep{BergerEtal2018, GoldbergBatygin2021}. In the Kepler-221 system, the radii of the planets are well constrained, but the masses of the planets are unknown due to weak TTVs \citep{BergerEtal2018, GoldbergBatygin2021}. In the system, planets b, c, and e are in a 6:3:1 3BR with the period ratios relatively far away from the integer period ratio (the period ratio between planets b and c is 2.035, and the period ratio between c and e is 3.228). However, planet d is not in any resonance with other planets. 

If the resonance chain in the Kepler-221 planetary system formed due to convergent migration during the disk phase, it is generally expected that the planets will be locked into a first-order resonance chain \citep{LeePeale2002}. However, except for planets b and c, this is not what is observed in the Kepler-221 system. Two observations stand out. First, planets c and e are close to (or in) 3:1 resonance, which is weaker than a first-order resonance. Second, planet d is clearly not in resonance with other planets (not even close to any). These two features cannot be explained by a simple disk migration model.

To explain the unique dynamical configuration of the Kepler-221 system, we propose a new dynamical model in which there were originally five planets in a first-order resonance chain, as would be expected from migration theory. The resonance breaks in the post-disk phase due to dynamical instability from infalling materials from a potential debris disk \citep{IzidoroEtal2017, IzidoroEtal2021, LiuEtal2022, NagpalEtal2024} or planetesimal scattering \citep{RaymondEtal2021, GhoshChatterjee2023, GriveaudEtal2024, WuEtal2024}. Therefore, the system became unstable, and two planets merged into the current planet d, while the three-body resonance between the other three planets (b, c, and e) reformed. Over long timescales, tidal dissipation fueled the expansion of these planets to their current period ratios. To investigate this scenario quantitatively, a modular approach was adopted, where we investigated the viability of each of these steps separately. The advantage of this approach is that the resulting conclusions are self-contained. In particular, when planet \bce{} 3BR expands toward the observed period ratios, we find that planet d puts additional constraints on the planet's tidal parameters and masses, which are independent of the detailed formation scenario.

This paper is structured as follows. In \se{model}, we present the dynamical configuration of the Kepler-221 system and the new four-phased dynamical model. In \se{successful}, we present a consistent simulation throughout all phases that successfully reproduced the observation configuration. Then we use N-body simulations to verify our model and conduct parameter studies following a chronological order. In \se{collision}, the evolution of the system up to the point of three-body reformation is discussed. \Se{expansion} investigates the orbital expansion phase, which puts constraints on the mass ratio of planets in resonance. According to the simulation results, we assess the five-planet formation model in \se{discussion}. Finally, we summarize our results and present them in \se{conclusion}.

\section{Model}
\label{sec:model}

\subsection{Resonance in the Kepler-221 planetary system}

Planets b, c, and e in Kepler-221 are very likely in resonance \citep{GoldbergBatygin2021}. The most prominent feature of two planets in resonance is that their period ratio is close to integer ratios. The more dynamically accurate description of resonance amounts to following the positions of the planet conjunctions in the orbit. This is encapsulated in terms of the resonance angle. The resonance angle can be expressed as:
\begin{equation}
	\phi_{1,2,X}=(j+o)l_1-jl_2-o\varpi_X,
	\label{eq:2BR_define}
\end{equation}
where $l_1$ and $l_2$ stand for the mean longitude for the inner and outer planet, $\varpi_X$  stands for either the longitude of periapsis for planet 1 ($X=1$) or planet 2 ($X=2$), which, respectively, correspond to the inner and outer resonance angle. In \eq{2BR_define}, $j$ and $o$ are integers called the resonance number and resonance order, respectively. For two-body resonances, first-order resonances are more stable ($o=1$) than higher-order resonances ($o>1$) \citep{MurrayEtal1999}. For two planets in resonance, either one or both inner and outer resonance angles should librate about a fixed value. For planets not in resonance, the resonance angle instead circulates from 0 to $2\pi$.

In a system with three or more planets, we can calculate the three-body resonance (3BR) angle by combining outer and inner two-body resonance angles and eliminating the longitude of periapsis of the middle planet. The expression of the 3BR angle is:
\begin{equation}
	\phi_{3BR}=p\lambda_1-r\lambda_2+q\lambda_3,
	\label{eq:3BR_define}
\end{equation}
where $\lambda_1$, $\lambda_2$, and $\lambda_3$ are the mean longitude for the three planets. The quantity $(p+q-r)$ is called the 3BR order. Planets are locked in 3BR if such an angle librates around a fixed value. The zeroth-order 3BR and first-order 3BR are generally stable and hard to break \citep{Petit2021}. For the Kepler-221 system, the period ratio between planets b and c is 2.035, and the period ratio between c and e is 3.228, which is close to 2:1 and 3:1, respectively. For simplicity, a librating $\phi_{3BR}=p\lambda_1-r\lambda_2+q\lambda_3$ is referred to by using the $(p,q,r)$ 3BR notation in the following. Therefore, planets b, c, and e might be in the $(2,3,5)$ zeroth-order 3BR. The B-value of a 3BR is defined by taking the time derivative of \eq{3BR_define}:
\begin{equation}
	B=\dot{\phi}_{3BR}=pn_1-rn_2+qn_3,
	\label{eq:B_define}
\end{equation}
where $n_1$, $n_2$, and $n_3$ are the mean motion for the three planets and $B$ is the B-value. For planets in 3BR, because their 3BR angle librates around a certain value, the B-value averaged over time should be close to zero.

The normalized B-value for a 3BR can be defined as  \citep{FabryckyEtal2014, GoldbergBatygin2021}:
\begin{equation}
	B_\mathrm{norm}=|B|/\langle n \rangle,
	\label{eq:B_norm}
\end{equation}
where $\langle n \rangle$ is the average of the mean motion of the planets. For planets b, c, and e of Kepler-221, $B_\mathrm{norm}=1.03\times 10^{-5}$ when $(p,q,r) = (2,3,5)$. Compared with other confirmed 3BR pairs such as K2-138 \citep{ChristiansenEtal2018}, Kepler-80 \citep{MacDonaldEtal2016}, and Kepler-60 \citep{GozdziewskiEtal2016}, the normalized B-value for Kepler-221 is statistically closer to zero, which implies that planets b, c, and e are in $(2,3,5)$ zeroth-order 3BR \citep{GoldbergBatygin2021}.

\subsection{The Formation of the Kepler-221 planetary system}
\label{sec:model_intro}

\begin{figure}
	\centering
	\includegraphics[width=\columnwidth]{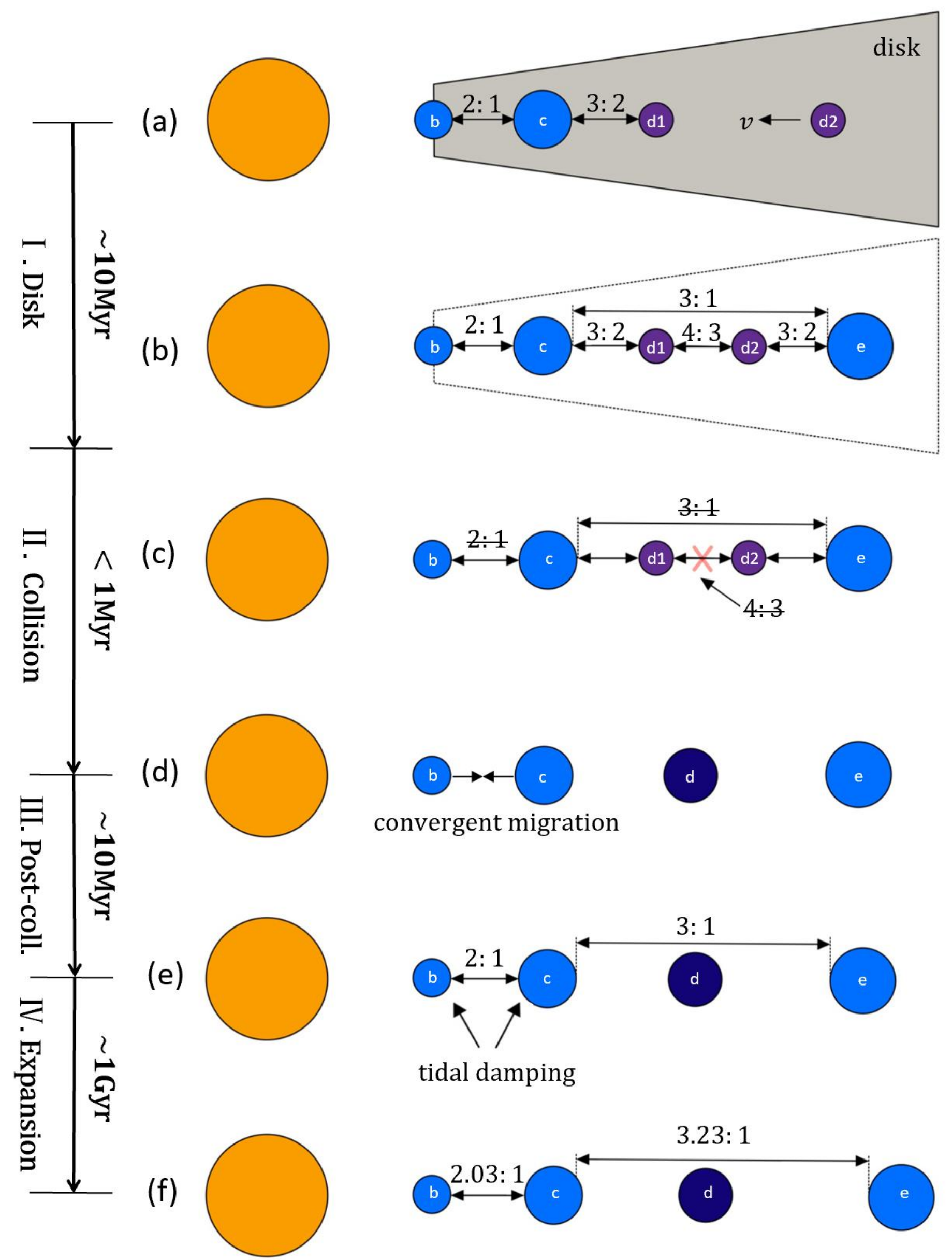}
	\caption{Schematic of the formation model for the Kepler-221 planet system investigated in this study. Sequential migration traps five planets in a chain of first-order resonances (panels a+b). After disk dispersal, the resonance chain breaks (panel c) triggering a dynamical instability that results in the merger of planets $\mathrm{d}_1$ and $\mathrm{d}_2$ (panel d). Convergent migration between planets b and c re-establishes the b, c, and e resonance chain (panel e). On evolutionary timescales (${\sim}1\,\mathrm{Gyr}$) the b, c, and e resonance chain expands to the present-day period ratios due to tidal dissipation (panel f).}
	\label{fig:model_sketch}
\end{figure}

To solve the puzzle of the formation of the \bce{} three-body resonance chain, we hypothesize that Kepler-221 originally harbored five planets, but two of them ($\mathrm{d}_1$ and $\mathrm{d}_2$) merged to form planet d. The model that we envision consists of four phases, as illustrated in \fg{model_sketch}. In chronological order:

\begin{enumerate}
	\item[I.]  In the \underline{disk phase}, five planets formed and then migrated inward due to Type-I migration. All planets are assumed to stay beyond the inner edge \citep{LiuEtal2017}. This leads to convergent migration and the adjacent planets are locked into a stable first-order two-body resonance chain: 2:1, 3:2, 4:3, and 3:2. The disk then disperses and the stable resonance chain remains, as shown in panel b in \fg{model_sketch}.
	\item[II.] In the \underline{collision phase}, planets $\mathrm{d}_1$ and $\mathrm{d}_2$ collide and merge into d. After the disk dispersal, instability first breaks the resonance between $\mathrm{d}_1$ and $\mathrm{d}_2$, which triggers system-wide instability that causes all resonances in the system to break (similar to what \citealt{RaymondEtal2021} investigate for the TRAPPIST-1 system). As planets $\mathrm{d}_1$ and $\mathrm{d}_2$ are the closest and lightest, they merge into one single planet d \citep{MatsumotoEtal2012}.

	\item[III.] In the third phase -- the \underline{post-collision} phase -- the \bce{} 3BR reforms (panels d and e of \fg{model_sketch}). Planets b, c, and e are initially not in resonance after the collision, but they stay very close to the zeroth-order 6:3:1 resonance position. Tidal damping in the system will damp the planets' eccentricities. Convergent migration due to tidal damping results in the reformation of the b, c, and e 3BR.

	\item[IV.] After reforming the zeroth-order 3BR, the system expands to the current period ratio -- the \underline{orbital expansion} phase (see panel f of \fg{model_sketch}). In this last phase, tidal damping will continue to expand the system while preserving the 3-body resonance  \citep{GoldbergBatygin2021}. This is because tidal damping preserves the angular momentum but dissipates the energy. As a result, the period ratio between planets in resonance in the system expands and the system reaches the observed position at present.
\end{enumerate}

In this scenario, the first three phases are essential to explain the presence of the 3-body \bce{} resonance chain in Kepler-221. The last phase explains the departure of period ratios from exact integer ratios through \bce{} 3BR expansion. This phase is independent of the details of the previous three phases; it only requires that planets b, c, and e are in resonance.

\begin{table}
	\caption{\label{tab:exp_para}Masses of the planets used in the simulation. }
	\centering
	\small
	\begin{tabular}{ccccc}
		\hline
		Mass model        & M1              & M2                                    & M2a & M2d\\
		                  & ``mass-radius'' & \multicolumn{3}{c}{``peas-in-a-pod''}        \\
		\hline
		$m_b$($m_\oplus$) & 5.09            & 5.09                                  & 5.09 & 5.09 \\
		$m_c$($m_\oplus$) & 9.08            & 5.09                                  & 6.50 & 6.50 \\
		$m_d$($m_\oplus$) & 8.52            & 5.09                                  & 5.09 & 10.18 \\
		$m_e$($m_\oplus$) & 11.55           & 5.09                                  & 5.09 & 5.09 \\ \hline
	\end{tabular}
\tablefoot{ 
In mass model M1, the masses of the planets follow the mass-radius relationship given by \citet{ChenKipping2017}. Mass model M2 assumes all planets have the same mass around $5M_\oplus$ consistent with the peas-in-a-pod mass model \citep{WeissEtal2020, WeissEtal2023}. Mass models M2a and M2d are variations of Model 2 following the mass constraint in \se{mass_con}. 
}
\end{table}

\begin{figure*}
	\centering
	\includegraphics[width=\textwidth]{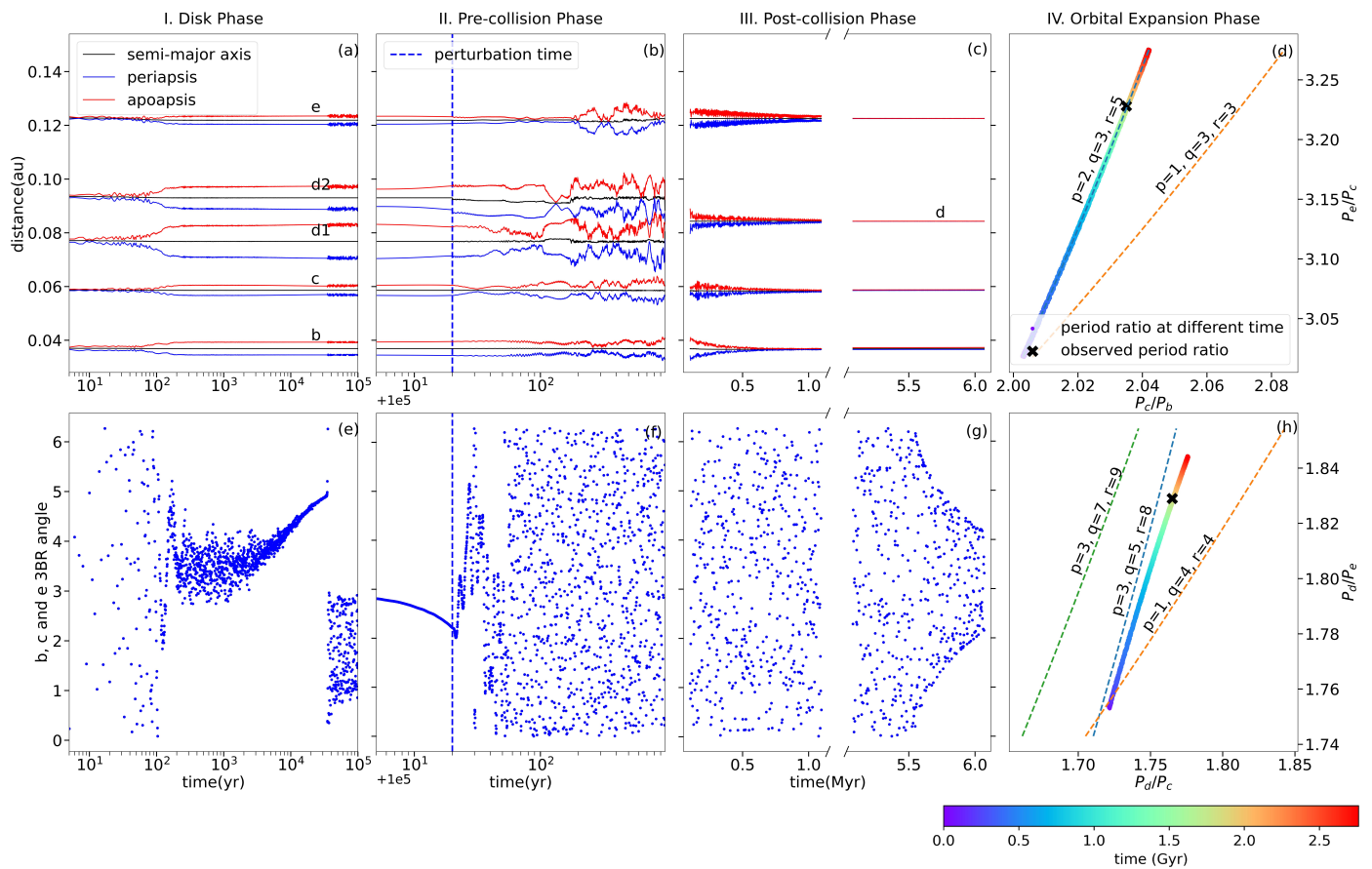}
        \caption{Successful simulation spanning all simulation phases depicted in \fg{model_sketch}. Panels a, b, and c show the semi-major axis and eccentricity evolution of the first three phases, and panels d, e, and f show the corresponding 3BR angle between planets b, c, and e. The evolution in the orbital expansion phase is represented by the period ratio evolution of planets c, d, and e in panels d and h.}
	\label{fig:all_8}
\end{figure*}

\subsection{Numerical implementation}
\label{sec:num_imp}
To verify the above scenario, we conduct a suite of N-body simulations using the \texttt{REBOUND} package \citep{ReinLiu2012}. We use the \texttt{WHFast} integrator \citep{ReinTamayo2015} in the disk phase with a timestep of 0.2\% of the orbital period of the innermost planet. In the collision, post-collision, and expansion phases, we use the \texttt{Mercurius} integrator \citep{ReinEtal2019}. The dissipation forces are implemented using the \texttt{REBOUNDx} package \citep{TamayoEtal2020}.

In the disk phase, planets migrate inward and resonances naturally form. The disk tidal force on a certain planet $i$ following \citep{PapaloizouLarwood2000} is:
\begin{equation}
	\bm{F}_\mathrm{disk,i}=-\frac{\mathbf{v}_\mathrm{i}}{2t_\mathrm{a,i}}-\frac{2\left(\mathbf{v}_\mathrm{i}\cdot \mathbf{r}_\mathrm{i}\right)\mathbf{v}_\mathrm{i}}{\left|r_\mathrm{i}\right|^2 t_\mathrm{e,i}},
	\label{eq:disk_force}
\end{equation}
where $t_\mathrm{a, i}$ and $t_\mathrm{e, i}$ correspond to the semi-major axis damping timescale and eccentricity damping timescale on planet $i$, respectively. After the disk has dispersed, stellar tides damp the planet eccentricities. While conserving angular momentum, the eccentricity of planets is damped on a timescale of \citep{PapaloizouEtal2018}:
\begin{equation}
	t_e=7.6\times 10^5Q_{\mathrm{phy}}\frac{m_i}{m_\oplus}\left(\frac{m_\odot}{m_\star}\right)^{1.5} \left(\frac{r_\oplus}{R_i}\right)^5 \left(\frac{a_i}{0.05\,\mathrm{au}}\right)^{6.5}\,\mathrm{yr},
	\label{eq:te_exp}
\end{equation}
where $m_i$, $a_i$ is the mass and semi-major axis of planet $i$ and $m_\star$ is the mass of the star. For simplicity, the simulations are conducted in the plane. 
In our simulations, we only apply eccentricity damping according to \eq{te_exp}. {However, it must be understood that the effective $Q_\mathrm{phy}$ accounts for additional damping mechanisms, such as obliquity tides. Therefore, $Q_\mathrm{phy}$ should not necessarily be identified with the tidal quality factor of a body. For example, when obliquity tides operate the equivalent tidal damping parameter, accounting for both tidal dissipation and obliquity tides, is expressed as  $Q_{\mathrm{phy}}^{-1}=Q_{\mathrm{d}}^{-1}+Q_{\mathrm{o}}^{-1}$, where $Q_\mathrm{o}$ is related to the obliquity $\epsilon$ of the planet (see \se{age_con}), $Q_{\mathrm{d}}=3Q/2k_2$, $Q$ is the tidal dissipation function of the planet, and $k_2$ is the Love number \citep{PapaloizouEtal2018}.

In addition, once planets are locked in 3BR in Phase IV, we accelerate the simulation by 100 times faster by adopting a $Q_\mathrm{sim}$ in the simulation that is much smaller than the effective $Q_{\mathrm{phy}}$: $Q_{\mathrm{sim}}=Q_{\mathrm{phy}}/100$. This will speed up the computations. A similar approach was used in previous literature \citep{PapaloizouEtal2018, HuangOrmel2022}.

The masses (see \tb{exp_para}) of the planets in the Kepler-221 system have not been directly inferred by observations due to weak TTV signals \citep{GoldbergBatygin2021}. Therefore, we work with three different mass models. In model 1 ``mass-radius'' it is assumed that the planets follow the mass-radius relationship derived by \citet{ChenKipping2017}. In model 2 ``peas-in-a-pod'' it is assumed that the planets are of equal mass \citep{WeissEtal2020, WeissEtal2023}. Finally, Model 2a and 2d are variations of Model 2 with a slightly more massive planet c and a twice as massive planet d.

\section{Charting the course: a successful simulation}
\label{sec:successful}

\begin{table*}[]
\renewcommand{\arraystretch}{1.2}
\centering
\caption{\label{tab:fig_para}Parameter setups in different simulation stages corresponding to figures from \fg{all_8} to \fg{orb_exp_suc}. }
\begin{tabular}{llllll}
\hline
Simulation stage        & Figures        & Phase(s)               & Mass model       & Parameter$^a$ & Value/Range              \\ \hline
\texttt{complete}       &\ref{fig:all_8}      & all         & M2a              & $m_\mathrm{d_2}/m_\mathrm{d_1}$       & $1.02$       \\
                           &                &                       &                 & $t_\mathrm{a,I}$       & $1.4\times 10^4\mathrm{yr}$              \\
                           &                &                       &                  & $t_\mathrm{a,I}/t_\mathrm{e,I}$ & $250$              \\
                           &                &                       &                  & $t_\mathrm{d}$ & $5\times 10^3\mathrm{yr}$              \\
                           &                &                       &                  & $\Delta a/a$ & $0.7\%$              \\
                            &                &                       &                  & $Q_\mathrm{phy,all}$  & $4.6$        \\
                           &                &                       &                  & $t^b_\mathrm{a0,III}$ & $1.9\times 10^8\mathrm{yr}$              \\
                           &                &                       &                  & $t^b_\mathrm{\Gamma,III}$ & $7.5\times 10^6\mathrm{yr}$              \\
\hline
\texttt{before planet merging} &\ref{fig:all_8} & I, II   & M2a, M2d  & $m_\mathrm{d_2}/m_\mathrm{d_1}$       & $[1,1.04]$       \\
                           &                &                       &                  & $t_\mathrm{a,I}$ & $[10^3,10^6]\mathrm{yr}$              \\
                           &                &                       &                  & $t_\mathrm{a,I}/t_\mathrm{e,I}$ & $[150,250]$              \\
                           &                &                       &                  & $t_\mathrm{d}$ & $[50,10^4]\mathrm{yr}$            \\
                           &                &                       &                  &  $\Delta a/a$ & $[0.5\%,1.2\%]$            \\
\hline
\texttt{3BR reformation} &\ref{fig:123_mech}, \ref{fig:tree_plot} & III   & M2a, M2d  & $Q_\mathrm{phy,c}$ & [3,10]      \\
                        &                &                       &                  & $Q_\mathrm{phy,c}/Q_\mathrm{phy,b}$ & $\left[0.01,1\right]$        \\
                        &                &                       &                  & $t^b_\mathrm{a0,III}$    &   $[6\times 10^7,2\times 10^8]\mathrm{yr}$         \\
                        &                &                       &                  & $t^b_\mathrm{\Gamma,III}$    &   $[2.5\times 10^6,7.5\times 10^6]\mathrm{yr}$         \\
\hline
\texttt{orbital expansion} &\ref{fig:all_8}, \ref{fig:orb_exp_fail}, \ref{fig:orb_exp_suc} & IV     & M1, M2a, M2d & $P_\mathrm{d}/P_\mathrm{c}$ & $[1.712,1.728]$ \\
                        &                &                       &                  & $t^b_\mathrm{a0,IV}$ &   $[4.6\times 10^9,2.3\times 10^{10}]\mathrm{yr}$         \\
                        &                &                       &                  & $t^b_\mathrm{\Gamma,IV}$ &   $[4.6\times 10^8,1.4\times 10^9]\mathrm{yr}$         \\
\hline

\end{tabular}
\tablefoot{$^a$Here the parameters denote -- $m_\mathrm{d2}/m_\mathrm{d1}$: the mass ratio between planet d$_\mathrm{2}$ and d$_\mathrm{1}$ before merging, $t_\mathrm{a,I}$: semi-major axis damping timescale in the disk phase, $t_\mathrm{a,I}/t_\mathrm{e,I}$: the ratio between semi-major axis damping and eccentricity damping timescale in disk phase, $t_\mathrm{d}$: the disk dispersal timescale, $\Delta a/a$: the instantaneous kick applied to the semi-major axis of planet d2 to break the resonances, $Q_\mathrm{phy,x}$: the effective tidal damping parameter for planet $x$ (see \se{num_imp}), $t^b_\mathrm{a0,III},t^b_\mathrm{a0,IV}$: the equivalent semi-major axis migration timescale due to torque on planet b in Phase III and Phase IV, $t^b_\mathrm{\Gamma,III},t^b_\mathrm{\Gamma,IV}$: the exponential decay timescale of the torque on planet b in Phase III and Phase IV, $P_\mathrm{d}/P_\mathrm{c}$: the period ratio of planet c and d after planet merging. }
\end{table*}

In this section, we prove the feasibility of the four-phase model by presenting a successful N-body simulation covering all four phases discussed in \se{model}. The successful simulation is shown in \fg{all_8} with the initial mass of planets according to model \texttt{M2a} in \tb{exp_para}, consistent with the mass constraint of the system (discussed in \se{allow_mass}). These and other parameters are discussed in detail in the later section; here we focus on the narrative. The general setup of the simulations in this work is listed in \tb{fig_para}.

The simulation starts with the disk phase as shown in panels a and e in \fg{all_8}. In the disk phase, five planets (planets b, c, $\mathrm{d}_1$, $\mathrm{d}_2$, and e) are initialized a little beyond the first-order resonance locations and semi-major axis damping and eccentricity damping are added on all planets except b, as shown in panel a. Planet b is assumed to stay at the disk inner edge, where positive co-rotation torques prevent planets from migrating further inward \citep{PaardekooperPapaloizou2009, LiuEtal2017}. This convergent migration locks the planets in a chain of first-order resonances around $100\,\mathrm{yr}$. The planets are quickly captured into two- and three-body resonances, including a 3BR between b, c, and e (panel e). The formation of resonances also excites the eccentricity of the planets (panel a). The eccentricities of planets after disk dispersal are determined by the disk damping parameters (See \se{phase12}).

After disk dispersal (end of Phase I), the planets in the simulation stay close to the resonance position (integer period ratio), which is typical for young resonant planetary systems \citep{HamerSchlaufman2024, DaiEtal2024}. Due to the adopted damping parameters the planets end up with a relatively high eccentricity of around 0.01-0.1 \citep{TeyssandierTerquem2014, PapaloizouEtal2018, YangLi2024}.

In the collision phase, dynamical instability is assumed to have taken place shortly after disk dispersal. The situation is illustrated in panels b and f of \fg{all_8}. After $t=20\,\mathrm{yr}$, a kick is applied to planet $\mathrm{d}_2$, which results in the breaking of the entire resonance chain. Because the breaking of the resonance chain allows for close encounters, the system becomes dynamically unstable which further excites the planet eccentricities. Because of the high (initial) eccentricities, orbit crossing follows rapidly \citep{ZhouEtal2007}, around $t=100\,\mathrm{yr}$ after resonance breaking, and merging of planets $\mathrm{d}_1$ and $\mathrm{d}_2$ is achieved after $800\,\mathrm{yr}$. 
Because of the short orbital instability phase, the other planets are still close to their resonant locations (see \se{phase12}). 
In the simulation, we assume, for simplicity, that two planets will experience perfect merging if their distance becomes closer to the sum of their radii. In reality, even if planets d1 and d2 did not have a perfect merger and experienced a hit-and-run collision, the “runner” is expected to return for a second giant impact resulting in planet merging \citep{AsphaugEtal2021}.

Panels c and g represent the post-collision phase in \fg{all_8}. After the merging of $\mathrm{d}_1$ and $\mathrm{d}_2$, the eccentricity of planet d decreases because planets $\mathrm{d}_1$ and $\mathrm{d}_2$ merged at their respective perihelion and aphelion distances, leading to a more circular orbit for planet d after merging \citep{KokuboIda1995}. Specifically, the eccentricity of planet $\mathrm{d}_1$ and $\mathrm{d}_2$ is around 0.1 before the merger while the eccentricity of planet d is around 0.05 after merging. Tidal damping also continuously decreases the eccentricity. In the post-collision phase, planets b and c undergo convergent migration due to the outward torque that is applied to planet b (see \se{orb_outmig}). Such an outward torque on planet b is essential to reform the resonance chain (see \se{P3-mec}). The 2:1 two-body resonance between planet b and c reforms first (after $1.7\,\mathrm{Myr}$; the two-body angle is not shown in \fg{all_8}). Thereafter, the c and e two-body 3:1 resonance and the b, c, and e 3BR reform simultaneously at around $5.7\,\mathrm{Myr}$. The conditions are now in place for the planet period ratios $P_\mathrm{c}/P_\mathrm{b}$ and $P_\mathrm{e}/P_\mathrm{c}$ to expand along the zeroth-order 3BR line (blue-dashed line in \fg{all_8}.

Panel d and h in \fg{all_8} show the evolution of the \bce{} and \cde{} period ratios during the orbital expansion phase (Phase IV). After the \bce{} 3BR reforms, tidal damping operates to expand the period ratio of the planets in resonance, migrating planets b and c inward and planet e outward. Initially, the planets are close to their corresponding 2:1 and 3:1 period ratios. The dashed lines in \fg{all_8} indicate different 3BRs characterized by $p$, $q$, and $r$ in \eq{B_define}:
\begin{equation}
	\frac{P_\mathrm{3}}{P_\mathrm{2}} = \left( \frac{r}{q} -\frac{pP_\mathrm{2}}{qP_\mathrm{1}} \right)^{-1},
	\label{eq:3BR_ratio}
\end{equation}
where $P_\mathrm{1}$, $P_\mathrm{2}$, $P_\mathrm{3}$ correspond to the inner, middle, and outer planets (i.e., \bce{} in \fg{all_8}d and \cde{} in \fg{all_8}h). In panel d of \fg{all_8}, tidal dissipation moves the planets along the blue dashed line corresponding to the $(2,3,5)$ zeroth-order 3BR between b, c, and e \citep{CharalambousEtal2018, PapaloizouEtal2018}. At the same time, the period ratio of planets c, d, and e also evolves along a line that avoids interaction with certain \cde{} 3BRs as shown in \fg{all_8}h. The slope of the expansion seen in \fg{all_8}h is related to the masses of the planets, as is discussed in \se{orb_exp_mec}. After about $2\,\mathrm{Gyr}$ of expansion, the period ratios of the planets are consistent with their observed values, implying a successful reproduction of the dynamical configuration of the Kepler-221 system.

In this section, a particular simulation has been selected to demonstrate the four-phase model. The successful result in \fg{all_8} is not guaranteed and the success rate of each phase is discussed in detail in \se{suc_rate}. Obviously, the successful outcome is somewhat tuned and parameter-dependent. Specifically, several key milestones can be identified. These include: the high eccentricity the planets acquire post-disk phase; the occurrence of a collision between planets d1 and d2; the reformation of the b/c two-body resonance, followed by the 3BR between \bce{}; the ability for sustained orbital expansion of the \bce{} system toward the current period ratios; and the inability of planet to dislodge these planets out of the 3BR they are in today. However, the successful result in \fg{all_8} serves the purpose of charting the course in this section. In the following, we discuss in detail the physical conditions that must be satisfied to pass each successive milestone and quantify the overall success rate of each model step.

\section{Early resonance formation, dynamical instability, and 3BR reformation (Phase I-III)}
\label{sec:collision}

In the previous section, we introduced a consistent simulation throughout all phases which successfully reproduced the observed configuration. In this section, the evolution from Phase I to III until the \bce{} resonance reformation will be investigated. The masses of planets in the simulations follow Model M2a in \tb{exp_para} in this section. The focus of this section will be on the conditions for which planet $\mathrm{d}_1$ and $\mathrm{d}_2$ manage to merge and the b, c, and e 3BR to successfully reform after the assumed planetary collision in Phase II. The range of parameters for the simulation from Phase I to III are listed in \tb{fig_para}.

\subsection{Evolution until dynamical instability}
\label{sec:phase12}

In the disk phase, the five planets (planets b, c, $\mathrm{d}_1$, $\mathrm{d}_2$, and e) form a stable first-order resonance chain by disk migration with planets b, c, and e in 6:3:1 resonance, as shown in panels a and e in \fg{all_8}. The mass of planet d is split between $\mathrm{d}_1$ and $\mathrm{d}_2$ such that its center of mass lies close to the observed position of planet d after the orbital expansion phase. Therefore the mass ratio of $m_{\mathrm{\mathrm{d}_1}}/m_{\mathrm{\mathrm{d}_2}}$ is assumed to be in the range of $[1,1.04]$. The formation of the first-order resonance chain is guaranteed because the planets are initialized close to their corresponding resonance position. To test the stability of the resonance chain after resonance formation, we apply a disk dispersal timescale $t_\mathrm{d}$ in the range of $[50\,\mathrm{yr}, 10^4\,\mathrm{yr}]$. The planets remain in resonance throughout disk dispersal. This is because the number of planets in the system is lower than the critical number in a first-order resonance chain \citep{MatsumotoOgihara2020}. Therefore, dynamical instability is required to break the resonance chain.  

It has been suggested that dynamical instability \citep{IzidoroEtal2017} or planetesimal scattering \citep{RaymondEtal2021} could perturb the system stability and break the resonance chain. Here, we do not model the specifics of the resonance breaking process, but simply enforce it by perturbing planet $\mathrm{d_2}$ with a "kick" just strong enough to break the $\mathrm{d}_1$ and $\mathrm{d}_2$ resonance. In the simulation, we apply a small instantaneous torque on planet $\mathrm{d_2}$ that decreases its semi-major axis ($\Delta a/a$) in the range of $[0.5\%,1.2\%]$ within $10^{-3} \mathrm{yr}$. The simulations show a $\Delta a/a$ as small as $0.7\%$ breaks the entire resonance chain \citep{ZhouEtal2007, PichierriEtal2020} and results in the rapid merging of planets $\mathrm{d}_1$ and $\mathrm{d}_2$, as shown in panels b and f in \fg{all_8}. 

In the collision phase, a quick merging between planets $\mathrm{d}_1$ and $\mathrm{d}_2$ is preferred. In that case, planets \bce{} would not experience much perturbation of their semi-major axes and would stay close to the 3BR resonance position, which is beneficial for reformation (see below). On the other hand, if the collision phase is long, the planets would undergo a long time of instability that would cause the positions of planets \bce{} to deviate significantly from the exact integer period ratio. This is disadvantageous for the subsequent 3BR reformation process. How rapid $\mathrm{d}_1$ and $\mathrm{d}_2$ merge after resonance breaking is determined by the eccentricity of planets ${\mathrm{d}_1}$ and ${\mathrm{d}_2}$. If the initial eccentricity is high (for instance, around 0.1), orbit crossing and merging of planets $\mathrm{d}_1$ and $\mathrm{d}_2$ will quickly follow the breaking of resonance. On the other hand, if the eccentricity remains low, the orbit of planets $\mathrm{d}_1$ and $\mathrm{d}_2$ would only cross after the planet eccentricity excites and the time required for the planet merging will be much longer \citep{TamayoEtal2021i}. Also, a higher eccentricity of planets in the resonance chains implies that they are deeper in resonance with the period ratio closer to integer \citep{HuangOrmel2023}. This is also advantageous for the 3BR reformation afterward with the initial position of the planet closer to the \bce{} 3BR position. The ratio between the semi-major axis damping timescale and eccentricity damping timescale $t_\mathrm{a, i}/t_\mathrm{e, i}$ determines the eccentricity of the planets after disk dispersal \citep{TeyssandierTerquem2014, PapaloizouEtal2018} with the eccentricity being proportional to the square-root of the damping ratio parameter. For example, in 45\% of the simulations planets d$_1$ and d$_2$ would merge within $10^3\,\mathrm{yr}$ with $t_\mathrm{a, i}/t_\mathrm{e, i}=250$ while the merging rate within $10^3\,\mathrm{yr}$ is lower at 10\% when $t_\mathrm{a, i}/t_\mathrm{e, i}=450$. Therefore, in 70\% of the simulations with $t_\mathrm{a, i}/t_\mathrm{e, i}=450$, planets \bce{} exhibit a large deviation from 3BR with $B_\mathrm{norm}>0.015$ after the merging of planets d$_1$ and d$_2$ (see \eq{B_norm}), making the 3BR reformation afterward almost impossible.

\begin{figure*}
	\centering
	\includegraphics[width=\textwidth]{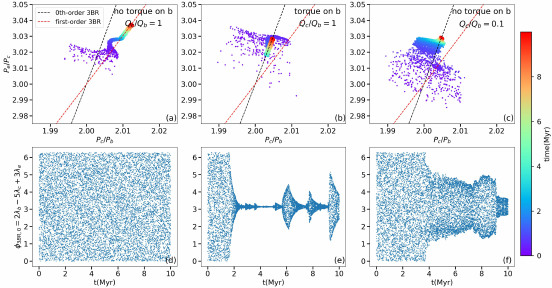}
	\caption{Success and failure in reforming the b, c, and e 3BR post-collision. Top panels show the evolution of $P_\mathrm{c}/P_\mathrm{b}$ and $P_\mathrm{e}/P_\mathrm{c}$. The black dashed line and red dashed line correspond to the zeroth-order and first-order 3BR, respectively, and the color of the dots indicates time. The bottom panels show the \bce{} 3BR angle. In panels a and d, there is no torque on planet b, and $Q_\mathrm{phy}$ is the same for all planets. The planets cross the zeroth-order 3BR and get trapped in a first-order 3BR, which is inconsistent with the present state. In panels b and e, planet b experienced an exponentially decaying outward torque with a timescale of $5\,\mathrm{Myr}$ with the same $Q_\mathrm{phy}$ for all planets. In panels c and f, tidal dissipation is more sufficient in planet c with $Q_\mathrm{c,phy}/Q_\mathrm{b,phy}=0.1$. }
	\label{fig:123_mech}
\end{figure*}
The success rate of mass model M2a and M2d (see \tb{exp_para} is tested in the simulation with the parameters in the range of \tb{fig_para}. With mass model M2a, in 51\% of the simulations planets d$_1$ and d$_2$ merge within $10^4 \mathrm{yr}$. In the other 49\% of the simulation, planets \bce{} already deviate a lot from the integer period ratio, making the future 3BR reformation impossible, so the simulation is truncated. On the other hand, the merging of other planets is possible when mass model M2d is assumed because of more massive planets d$_1$ and d$_2$. In 47\% of the simulations still d$_1$ and d$_2$ merge within $10^4\,\mathrm{yr}$. In 6\% of the simulations planets d$_2$ and e merge while in 11\% of the simulation planets c and d$_1$ merge. The other simulations are truncated since planets \bce{} already deviate from the 3BR position.

\subsection{Resonance reformation after dynamical Instability}
\label{sec:phase3}

Panels c and f of \fg{all_8} represent the post-collision phase after the merging of planets d$_1$ and d$_2$. While the success rate in Phase I and II is relatively high, the reformation of the \bce{} 3BR after dynamical instability is much more challenging. \Fg{all_8} features in fact a successful condition in which the b, c, and e 3BR reforms because of an outward (positive) torque applied on planet b. This ensures that planets b and c convergently approach each other, facilitating 2BR formation (see \se{P3-mec}). Otherwise, the reformation of the (2, 3, 5) zeroth-order 3BR cannot be accomplished \citep{Petit2021}. Specifically, the desired scenario is that planets b and c first reform the 2:1 two-body resonance under convergent migration, which is relatively straightforward. Then b, c, and e 3BR form at the same time with the c and e in 3:1 resonance. However, in the scenario where $Q_\mathrm{phy}$ is the same for all planets, planets b and c tend to experience divergent migration. As a result, there is only a slight chance to reform the b, c, and e 3BR.

We run over 50 simulations with the setup applying the same $Q_\mathrm{phy}$ (see \se{num_imp}) on all planets but with slightly different initial positions and eccentricities of the planets for each simulation. In all these simulations, planets b, c, and e fail to reform the 3BR. An illustrative example is given by panels a and d in \fg{123_mech}. Instead of forming the desired 3BR, planet \bce{} either form a first-order 3BR (40\%; as in \fg{123_mech}a) or the formation of the 3BR fails altogether (60\%). The reason why a first-order 3BR forms but not the zeroth-order is because the formation of a first-order 3BR is not preconditioned on the requirement of being close to exact resonance (see \App{1st3BR}) \citep{Petit2021}. Therefore, in the absence of convergent motions, the planets would either fail to be locked in resonance or form the first-order 3BR and expand along it (red dashed line in \fg{123_mech}).

\subsection{Potential mechanisms facilitating resonance reformation}
\label{sec:P3-mec}

In the previous paragraph, we outlined that convergent migration between planets b and c is essential to reform the two-body and three-body resonances between planets b, c, and e. Here, we envision two such mechanisms that can decrease $P_\mathrm{c}/P_\mathrm{b}$: i) a positive torque on planet b; ii) an enhanced tidal dissipation on planet c.

The first mechanism is a potential outward torque operating on planet b, which promotes the convergent migration between b and c and reforms the resonance. Panels b and e in \fg{123_mech} show the 3BR reformation process corresponding to an exponentially decaying outward torque on planet b. The origin of the torque could include mass loss on planet b \citep{WangLin2023} or dynamical tides \citep{AhuirEtal2021}. In the simulation, we take the form of the exponentially decaying torque:
\begin{equation}
	\Gamma = \Gamma_{\mathrm{0}} \exp{\left(-\frac{t}{t_\Gamma}\right)},
	\label{eq:gamma_4}
\end{equation}
where $t$ is the simulation time after the merging of planets d1 and d2, $\Gamma_0$ is the value of the torque at $t=0$ and $t_\Gamma$ is the timescale on which the torque decays. The initial value of the torque can equivalently be parameterized in terms of an orbital decay timescale, for instance, if it acts on planet b:
\begin{equation}
    t_{a0}^\mathrm{b} \equiv \frac{L_\mathrm{b}}{2\Gamma_0},
\end{equation}
where $L_b$ is the current total angular momentum of planet b. This is the form given in \tb{fig_para}.

In panel c and g of \fg{all_8}, the torque applied to planet b amounts to a 2\% mass loss in $7.5\,\mathrm{Myr}$ during the post-collision phase, which corresponds to an initial torque of $\Gamma_{\mathrm{0}}=8\times 10^{24}\,\mathrm{N\cdot m}$. Expressed in terms of the parameters defined above, $t^\mathrm{b}_\mathrm{a0,III}=2\times 10^8\,\mathrm{yr}$ and $t^\mathrm{b}_\mathrm{\Gamma,III}=7.5\,\mathrm{Myr}$. Mass loss of planet b would also excite the eccentricity of the planet on a timescale of $t_\mathrm{e}=2m/\gamma\dot{m}$ \citep{CorreiaEtal2020}. At the stated mass loss rates the corresponding eccentricity excitation timescale therefore amounts to a minimum eccentricity excitation timescale (with $\gamma=1$) of $t_\mathrm{e}=10^8\,\mathrm{yr}$ (such timescale is even longer in Phase IV, see \se{orb_outmig}). Because tidal damping also operates on the planets with a timescale of $10^6\,\mathrm{yr}$ (see \eq{te_exp}), it is safe to neglect the excitation of eccentricity due to mass loss.

Panel b of \fg{123_mech} presents the period ratio diagram of a simulation with the imposed torque operating on planet b. It can be seen that the planet period ratio first oscillates around the resonance position relatively far from the integer period ratio. Then the positive torque on b decreases $P_\mathrm{c}/P_\mathrm{b}$ while quickly trapping planets b and c in two-body resonance. Planets b, c, and e also quickly get captured in 3BR, as can be seen from the librating resonance angle shown in panel e of \fg{123_mech}. Subsequently, the period ratio of the planets expands along the zeroth-order resonance line.

The second mechanism to promote resonance reformation is more efficient damping on planet c. Planet c could potentially have a relatively small $Q_\mathrm{phy}$ (with $Q_\mathrm{c,phy}/Q_\mathrm{b,phy}=0.1$) due to efficient obliquity tides \citep{LoudenEtal2023, GoldbergBatygin2021} while the effective tidal $Q$ parameter for other planets is much larger (see \se{age_con}). Tidal damping on planet c would result in its inward migration and contribute to the convergent migration between planets b and c, necessary to the formation of the b and c 2BR and the b, c, and e 3BR. Panel c and f in \fg{123_mech} correspond to such a simulation. Here, damping on planet c is stronger, with a smaller tidal $Q$ parameter for planet c, ($Q_\mathrm{c}/Q_\mathrm{b}=0.1$). As shown in the figure, planets b, c, and e first oscillate around the zeroth-order resonance (black dashed line) with large amplitude. Then tidal damping on the system migrates planets b and c inward, decreasing $P_\mathrm{c}/P_\mathrm{b}$ while increasing $P_\mathrm{e}/P_\mathrm{c}$. The subsequent convergent migration first forms the b and c 2BR. Afterward, tidal damping on planets b and c migrates c outward, leading to the convergent migration between planets c and e and the reformation of \bce{} 3BR.

The reformation of the \bce{} 3BR requires convergent migration between planets b and c. Otherwise, planets b, c, and e will be trapped in the first-order 3BR or will not get trapped in any 3BR. Here, we have demonstrated the viability of resonance reformation for two physically plausible scenarios of an outward torque on planet b and more effective tides on planet c. Nevertheless, the reformation of the \bce{} 3BR is stochastic and cannot be guaranteed. This is discussed in more detail in \se{suc_rate}.

\begin{figure}
	\centering
    \includegraphics[width=\columnwidth]{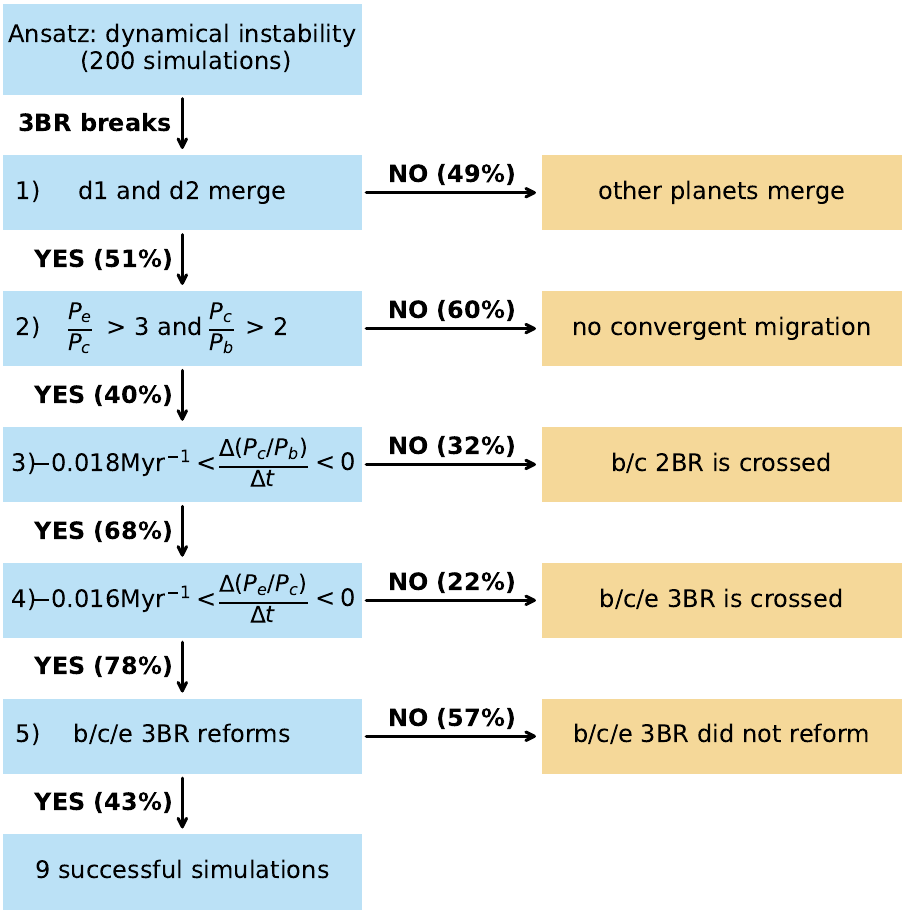}
	\caption{\label{fig:tree_plot}Tree plot showing the outcome and success rate of simulations in the post-collision phase. The total success rate of the simulation in the post-collision phase is 9\% (9 success in 101 simulations) but conditional success probabilities are much higher.}
\end{figure}

\subsection{Success rate analysis}
\label{sec:suc_rate}

To understand the success of resonance reformation post-collision, we run 200 simulations with different parameters (see \tb{fig_para}). In 101 of them planets $\mathrm{d_1}$ and $\mathrm{d_2}$ collide, as shown in \fg{tree_plot}, while in the other simulations planets c and $\mathrm{d_1}$ collide, or some planets are ejected out of the system. The overall success rate in the post-collision phase is around 10\% (9 successful in 101 simulations). In the simulation, an outward torque is applied on planet b corresponding to a mass loss of 2\% to promote the convergent migration between planets. Tidal damping is applied on all planets with the tidal $Q_\mathrm{phy}$ parameter the same on each planet and varies from 3 to 10. A simulation is categorized as a successful simulation if planet \bce{} successfully gets trapped in the zeroth-order 3BR. 

\begin{figure*}
	\sidecaption
	\centering
	\includegraphics[width=0.7\textwidth]{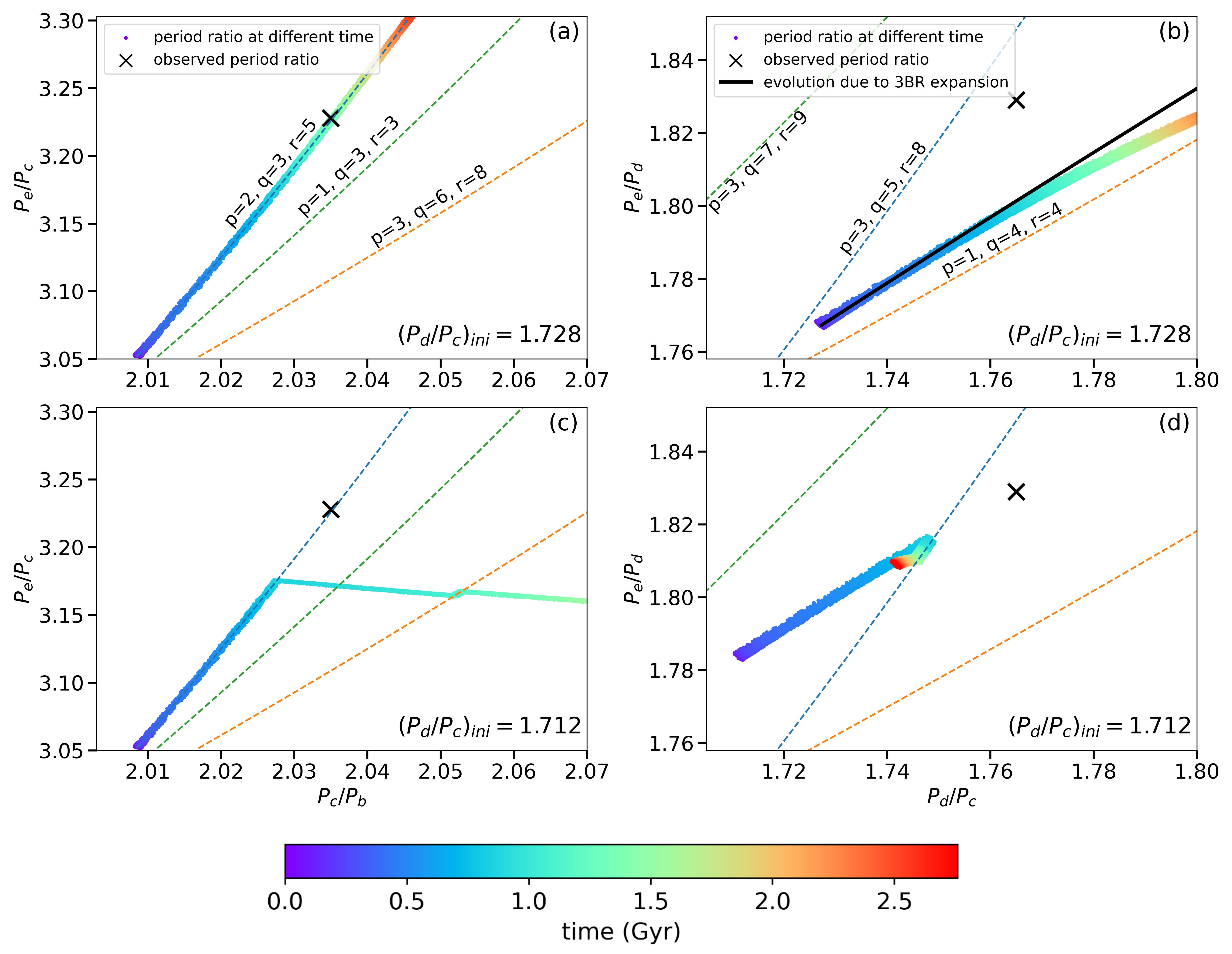}
	\caption{\label{fig:orb_exp_fail}Effect of the initial location of planet d on the expansion process of the Kepler-221 system. Two simulations with identical masses of the planet following mass model \texttt{M1} (\tb{exp_para}) and different initial positions are shown from top to bottom. The left panels show the evolution of $P_\mathrm{c}/P_\mathrm{b}$ and $P_\mathrm{e}/P_\mathrm{c}$ and the right panels show the evolution of $P_\mathrm{d}/P_\mathrm{c}$ and $P_\mathrm{e}/P_\mathrm{d}$. In the panels, colored dots represent the evolution of the period ratios with the color bar indicating time. The black crosses represent the current period ratio of the system. Dashed lines of different colors correspond to different 3BRs between b, c, and e or c, d, and e, with $p$, $q$, and $r$ defined in \eq{3BR_define}. The black line in panel b represents the analytical evolution of the period ratio only considering b, c, and e 3BR expansion with the position of d fixed.}
\end{figure*}

The resonance reformation process is a critical step of the model. To obtain meaningful statistics on its viability we conduct a parameter study of Phases II and III. Altogether 200 simulations are run varying parameters such as tidal damping strength and additional torque on planet b. As it is already shown that convergent migration is critical, an outward torque is applied on planet b with $t^\mathrm{b}_\mathrm{0,III}$ between $[2.5,7.5]\,\mathrm{Myr}$ and $t^\mathrm{b}_\mathrm{a,III}$ in the range of $[6\times 10^7,2\times 10^8]\mathrm{yr}$ (see above, \se{P3-mec}). Tidal damping is applied on all planets with the tidal $Q_\mathrm{phy}$ parameter the same for each planet and varying from 3 to 10. A simulation is categorized as successful if planet \bce{} successfully gets trapped in the zeroth-order 3BR. Based on these simulations, we identify milestones that must be taken to guarantee resonance reformation. The results are presented graphically in \fg{tree_plot} in the form of a decision tree. Although the bulk success rate is low, less than 10\%, \fg{tree_plot} shows that it can be significant when certain conditions and Ansatzes are met.

Reformation of the 3BR is only possible when $P_\mathrm{c}/P_\mathrm{b}>2$ and $P_\mathrm{e}/P_\mathrm{c}>3$ immediately post-collision (this happens in 40\% of the simulations). Here, the period ratio is measured as the average in the first $100\,\mathrm{yr}$ after merging of planets $\mathrm{d_1}$ and $\mathrm{d_2}$. Period ratios above exact resonance are needed in order to set up conditions for convergently migrating planets into resonance (see \se{phase3}). In 68\% of these simulations, planets b and c undergo a relatively slow convergent migration ($-0.018\mathrm{Myr}^{-1}<\Delta(P_\mathrm{c}/P_\mathrm{b})/\Delta t<0$) and the b/c 2BR successfully reforms. Due to the outward torque on b and the tidal damping on the planets, planet c would migrate outward and undergo convergent migration with planet e after the formation of b/c 2BR. The trapping of planets \bce{} prefer slow migration of c (with $-0.015\mathrm{Myr}^{-1}<\Delta(P_\mathrm{e}/P_\mathrm{c})/\Delta t<0$, occurring in 78\% of the simulation). Otherwise, if $P_\mathrm{e}/P_\mathrm{c}$ decreases too rapidly, the period ratio between planets c and e will quickly drop below three, and the trapping of c/e 2BR and \bce{} 3BR becomes impossible. When all these migration criteria are satisfied, the \bce{} 3BR has a chance of forming in 43\% of the simulation (9 out of 21).

In reality, the situation could belong in this category because we omitted the higher $Q_\mathrm{phy}$ parameters from our parameter study due to computational constraints. Specifically, we have enforced a maximum simulation time of $10\,\mathrm{Myr}$, within which the resonance reformation must take place. Therefore, a more gentle convergent migration during the formation of b/c 2BR and \bce{} 3BR would naturally satisfy conditions 3) and 4) in \fg{tree_plot} with a higher tidal Q parameter \citep{Wu2005, JacksonEtal2008}. On the other hand, it is impossible to reform the \bce{} 3BR with $Q_\mathrm{phy}<3$ due to too rapid migration. 
Therefore, given the Ansatz that a collision took place, the (maximum) success rate of the \bce{} resonance reformation stands at about 17\% (=$0.4{\times}0.43$).

Finally, we run a similar success rate analysis for mass model M2d in \tb{exp_para} assuming planet d is about twice as massive as other planets. The success rate of 3BR reformation in Phase III is around 10\%, similar to mass model M2a. This implies that a massive planet d, potentially due to the merging of planets d1 and d2, has minimal impact on the reformation of \bce{} 3BR in Phase III. In conclusion, the reformation of 3BR in the post-collision phase appears viable under plausible physical conditions.

\section{Orbital expansion (Phase IV)}
\label{sec:expansion}

After the formation of \bce{} 3BR, discussed in \se{collision}, the period ratios between b, c, and e are still close to integer. On the other hand, planets b, c, and e in the Kepler-221 system are currently locked into three-body resonance, relatively far away from integer period ratio, with $P_\mathrm{c}/P_\mathrm{b}=2.035$ and $P_\mathrm{e}/P_\mathrm{c}=3.228$. This implies that Phase IV -- the orbital expansion phase -- must have happened in the history of the Kepler-221 system \citep{GoldbergBatygin2021}, regardless of the formation origin of the 3BR. The 3BR expansion excluding planet d in the Kepler-221 system was already investigated in \citet{GoldbergBatygin2021}. In this section, we run the expansion phase including planet d, and demonstrate under which conditions tidal damping acting on the b, c, and e three-body resonance is capable of expanding these planets to their current observed period ratios. We will show that planet d can significantly affect the orbital expansion phase due to \bcd{} 3BR crossing, which could put constraints on the system parameters including the mass ratio of planets.

\subsection{Orbital expansion: Failure and success}
\label{sec:orb_exp_mec}

\begin{figure*}
	\sidecaption
	\centering
	\includegraphics[width=0.7\textwidth]{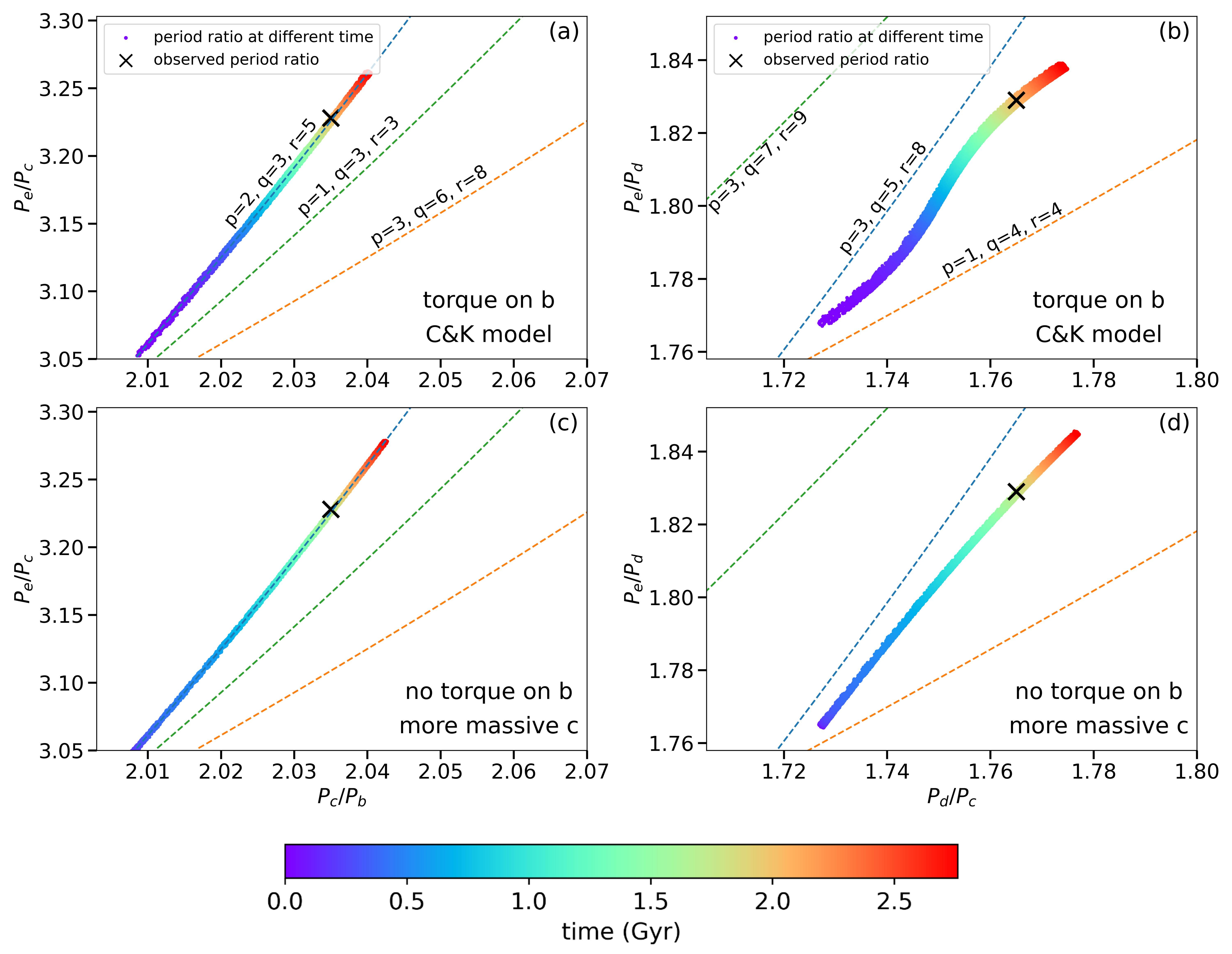}
	\caption{\label{fig:orb_exp_suc}Mechanisms that ensure successful orbital expansion of the Kepler-221 system. Two simulations with identical initial positions of the planets and different tidal strengths, torque on planet b, and planet masses are shown from top to bottom. (i) Top panels: A temporal outward torque on planet b with masses according to \texttt{M1} in \tb{exp_para}; (ii) Bottom panels: different initial masses of planets according to \texttt{M2a} in \tb{exp_para}.}
\end{figure*}

In the orbital expansion phase, a successful simulation matching the observed period ratio has two requirements: i) planets \bce{} should stay in resonance and expand to the current period ratio $P_\mathrm{c}/P_\mathrm{b}=2.035$ and $P_\mathrm{e}/P_\mathrm{c}=3.228$; ii) when planets \bce{} reach the current period ratio, planet d should also match the observed period ratio with $P_\mathrm{d}/P_\mathrm{c}=1.765$ and $P_\mathrm{e}/P_\mathrm{d}=1.829$. A successful expansion matching these two conditions, as in panels d and g in \fg{all_8}, is not guaranteed.

Because the orbital expansion phase is not related to the formation process of \bce{} 3BR, we used an optimized model to form the \bce{} 3BR to simplify the parameter setup as the initial masses of the planets. We initialize five planets (b, c, $\mathrm{d_1}$, $\mathrm{d_2}$ and e) in the first-order resonance chain consistent with Phase I and adiabatically decrease the masses of $\mathrm{d_1}$ and $\mathrm{d_2}$ to zero to ensure that the remaining planet b, c and e are still in 3BR close to integer period ratio. Then we add planet d back. The detailed setup of the optimized model is discussed in \App{opt_model}.

\Fg{orb_exp_fail} shows the outcome of two unsuccessful simulations with the same masses of planets following \texttt{M1} in \tb{exp_para} and different assumptions on the initial value of $P_\mathrm{d}/P_\mathrm{c}$. Initially, the planets are close to their corresponding 2:1 and 3:1 period ratios, which is a condition for the trapping in the zeroth-order 3BR (see \se{phase3}). The dashed lines in \fg{orb_exp_fail} indicate different 3BRs characterized by $p$, $q$ and $r$ in \eq{3BR_ratio}. Tidal dissipation moves planets along $(2,3,5)$ zeroth-order 3BR between b, c, and e \citep{CharalambousEtal2018, PapaloizouEtal2018}, indicated in the left panels of \fg{orb_exp_fail} by the blue dashed line.

While the \bce{} system starts from near-exact resonance locations, there is some freedom to choose the initial starting point of planet d. In \fg{orb_exp_fail}, panels a and b present a simulation where planet d is positioned below the $(p,q,r)=(3,5,8)$ 3BR resonance line. In addition, only tidal damping operates on the planets and all planets have their $Q_\mathrm{phys}=4.6$, which is consistent with the successful simulation in \fg{all_8}. These low Q-values are solely adopted for computational expediency since they speed up the simulation. However, as is discussed in more detail in \se{age_con} the young age demands a very efficient damping mechanism. The adopted value for $Q_\mathrm{phys}$ in the orbital expansion phase simulations does not affect the simulation outcome. A larger $Q_\mathrm{phys}$ will not change the simulation outcome, but it will make the duration of the orbital expansion to the observed period ratio longer. Therefore, the age of the system (younger than $650\;\mathrm{Myr}$) might be inconsistent with the simulation, as explained in \se{age_con}. During tidal expansion the periods of planets b, c, and e change, following (by definition) the $(2,3,5)$ 3BR line (panel a), while $P_\mathrm{d}$ remains largely constant. As a result, the increase of $P_\mathrm{e}$ and decrease of $P_\mathrm{c}$ lead to the increase of both $P_\mathrm{e}/P_\mathrm{d}$ and $P_\mathrm{d}/P_\mathrm{c}$ and the system moves away from the (3,5,8) resonance line (panel b). Therefore, during the orbital expansion planets c, d, and e would not encounter this 3BR, which could dislodge it from the \bce{} resonance (see below), and planets b, c, and e successfully expand to the observed period ratio (panel a). Nevertheless, this scenario fails as the present location of planet d with respect to c an e (the black cross in \fg{orb_exp_fail}b) cannot be reached.

The slope in the $P_\mathrm{e}/P_\mathrm{d}$ versus $P_\mathrm{d}/P_\mathrm{c}$ plane (\fg{orb_exp_fail}b) is the result of the conservation of the 3BR resonance angle and angular momentum. Specifically, without considering the potentially disturbing effects of planet d, the slope in this plane can be analytically solved (see \App{cde-slope}):
\begin{equation}
	a_\mathrm{cde}=\frac{\Delta({P_\mathrm{e}/P_\mathrm{d}})}{\Delta(P_\mathrm{d}/P_\mathrm{c})}
	= \frac{4.90m_\mathrm{b}+4.94m_\mathrm{c}}{7.27m_\mathrm{e}-m_\mathrm{b}},
	\label{eq:slope_for}
\end{equation}
From \eq{slope_for}, it is obvious that the slope of the expansion increases with a more massive planet b and c or a less massive e. In \fg{orb_exp_fail}b, the effect of the b, c, and e 3BR expansion on the evolution of $P_\mathrm{d}/P_\mathrm{c}$ and $P_\mathrm{e}/P_\mathrm{d}$ is represented by the black line, which has a slope of 0.89. The simulation including planet d, represented by the colored line in panel b in \fg{orb_exp_fail}, shows little difference from the analytical solution represented by the black line. The difference stems from a minor outward migration of planet d due to secular interactions with the other planets (see \App{d-mig}). Therefore, the value of $P_\mathrm{d}/P_\mathrm{c}$ in the simulation (colored line) is a little larger with respect to the analytical prediction (black line).

A possible solution to match the low $a_\mathrm{cde}$ is to decrease the initial semi-major axis of planet d, and thereby $P_\mathrm{d}/P_\mathrm{c}$, in such a way that it could reach the correct observed position. One example is shown in panels c and d of \fg{orb_exp_fail}. However, this simulation is also unsuccessful, because planets c, d, and e would encounter the $(3,5,8)$ 3BR during the orbital expansion, as shown in panel d. The resonance overlapping experienced at this point would break the b, c, and e 3BR and stop the orbital expansion of these planets \citep{PetitEtal2020i}. We find that the \bce{} 3BR are dislodged in more than 95\% simulations, similar to what happened in panel c.

Therefore, the system likely started its orbital expansion right of the $(3,5,8)$ 3BR line in the \ppp{e}{d}{c} period plane. Reaching the observed period ratio is only possible with:
\begin{equation}
	a_\mathrm{cde} > 1.56,
	\label{eq:slope_for2}
\end{equation}
Thus, the expansion of $P_\mathrm{d}/P_\mathrm{c}$ must feature some degree of slowdown, while that of $P_\mathrm{e}/P_\mathrm{d}$ must not. We envision two possible mechanisms to achieve this, which are shown in \fg{orb_exp_suc}.

\begin{enumerate}
	\item A (temporary) slowdown in the expansion of $P_\mathrm{d}/P_\mathrm{c}$ due to outward migration of planet c, for example through a positive torque acting on planet b.  Because planets b, c, and e are linked by 3BR, the outward migration of b would also lead to the increase in the semi-major axis of c and e. An example of a scenario that includes positive torques is shown in panels a and b of \fg{orb_exp_suc}. Specifically, a positive torque, with $t_\mathrm{a0,IV}^\mathrm{b}=2.1\times 10^{10}\,\mathrm{yr}$ and $t_\Gamma^\mathrm{b}=1.4\,\mathrm{Gyr}$, is added to planet b (see below, \se{orb_outmig}). The torque slows down the expansion of $P_\mathrm{d}/P_\mathrm{c}$ with respect to $P_\mathrm{e}/P_\mathrm{d}$. Its decaying nature results in an S-shaped trajectory in the \cde{} period ratio plane (\fg{orb_exp_fail}b). As a result, the planets successfully reach the observed period ratio near the end of the simulation (see \se{orb_outmig}).

	\item A combination of planet masses that results in sufficiently steep $a_\mathrm{cde}$. A corresponding simulation is shown in panels c and d of \fg{orb_exp_suc}. As shown in \eq{slope_for}, increasing the slope in the \ppp{e}{d}{c} plane can be achieved with a more massive b or c or a less massive e.  An example is mass model \texttt{M2a} in \tb{exp_para}, which assumes a slightly more massive planet c with respect to the other planets but is still in line with the peas-in-a-pod mass model \citep{WeissEtal2020, WeissEtal2023}.  The orbital expansion resulting from these planet masses is shown in panels e and f in \tb{exp_para}. Due to the increased slope in the $P_\mathrm{e}/P_\mathrm{d}{-}P_\mathrm{d}/P_\mathrm{c}$ plane, the initial position of planet d can be chosen to avoid the $(3,5,8)$ 3BR. This scenario also ensures a successful orbital expansion to the observed period ratio.
\end{enumerate}

\citet{GoldbergBatygin2021} expanded b, c, and e 3BR, but planet d was neglected because its current position is far from a 3BR with any other planet. We stress, however, that planet d is crucial in the orbital expansion phase. Although planet d is far from 3BR currently, when planets c, d, and e hit a 3BR during the orbital evolution, it could break the 3BR between b, c, and e and result in a different architecture from the observation. Therefore, it is necessary to include planet d in the orbital expansion simulation. The initial, relative positions of planet d and the \cde{} 3BR are crucial for the subsequent orbital expansion of the system.

\subsection{Orbital expansion: Specific scenarios}
\label{sec:orb_parstudy}
In this subsection, we conduct parameter studies for the two mechanisms outlined above: (i) outward migration of the \bce{} system; or (ii) a change in the masses of the planets.

\subsubsection{Outward-moving planet b}
\label{sec:orb_outmig}
\begin{figure}
	\includegraphics[width=\columnwidth]{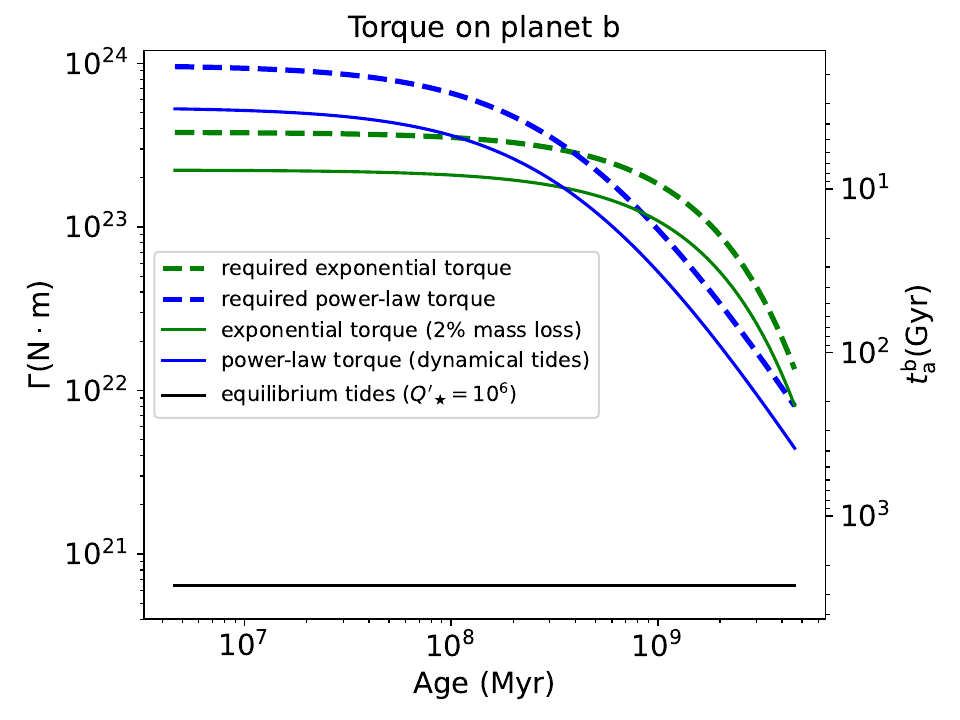}
	\caption{\label{fig:toeque_b}Time dependence of torque on b for different physical mechanisms. The dashed lines represent the minimum dynamical torque on (blue) or mass loss rate for (green) planet b required to induce a high enough migration of planet c, which ensures sustained orbital expansion of the \bce{} system. These torques decay with time according to the models by \citet{WangLin2023} (mass loss) and \citet{AhuirEtal2021} (dynamical torques). For reference, the green solid line represents the torque on b due to a 2\% mass loss \citep{WangLin2023} and the blue solid line fits the dynamical tide model of \citet{AhuirEtal2021}. The black line represents the torque due to equilibrium tides assuming $Q_\star'=10^6$.}
\end{figure}

The first mechanism is the outward migration of planet b. If the masses of the planets follow \texttt{M1} in \tb{exp_para}, the planets could not reach the observed period ratio, as shown in \fg{orb_exp_fail}. However, the planets could be secured into the observed position if the outward migration of planet b slows down the increase of $P_\mathrm{d}/P_\mathrm{c}$. The origin of the torque could be due to multiple reasons, including mass loss \citep{Carroll-NellenbackEtal2017, WangLin2023, VazanEtal2024} or dynamical tides \citep{BolmontMathis2016, BenbakouraEtal2019, AhuirEtal2021}. Importantly, these torques are time-dependent and operate only at early times. 

For isotropic mass loss, the total angular momentum is conserved and the planet migrates to a higher orbit upon losing mass. This process effectively amounts to a torque operating on the planet. If the mass loss fraction is $A$ and the exponential decaying mass loss timescale is $\tau_\mathrm{m}$, the corresponding torque on planet b according to \citet{WangLin2023}, $\Gamma_\mathrm{ML}$ is
\begin{equation}
	\Gamma_\mathrm{ML} = \frac{L_\mathrm{b}}{\tau_\mathrm{m}}\frac{A\exp(-t/\tau_\mathrm{m})}{1+A\exp(-t/\tau_\mathrm{m})} \approx \Gamma_\mathrm{ML,0} \exp (-t/\tau_\mathrm{m}),
\end{equation}
where $\Gamma_\mathrm{ML,0}=AL_\mathrm{b}/\tau_\mathrm{m}$, $t$ is the simulation time after the 3BR reformation, $L_\mathrm{b}$ is the current total angular momentum of planet b. Such torque is equivalent to a semi-major axis expansion timescale $t^b_\mathrm{a0,IV}$ following \eq{gamma_4} with $t^b_\mathrm{\Gamma,IV}=\tau_\mathrm{m}$. In the orbital expansion phase, if the masses of the planets follow model \texttt{M1} in \tb{exp_para}, the torque required to expand the planet to the observed position corresponds to $t^b_\mathrm{a0,IV}=2.1\times 10^{10}\mathrm{yr}$ and $t^b_\mathrm{\Gamma,IV}=1.4\,\mathrm{Gyr}$. Notice that $t^b_\mathrm{\Gamma,IV}$ is much longer than the corresponding timescale in Phase III ($t^b_\mathrm{\Gamma,III}$, see \tb{fig_para}) because the mass loss operates over a longer timescale in Phase IV.

On the other hand, dynamical tides are better described with a power-law decay model. We fit the results obtained by \citet{AhuirEtal2021}, where a rapid decline of the torque is seen to occur after a time $\tau_\mathrm{d}$:
\begin{equation}
	\Gamma_\mathrm{DT} = \Gamma_\mathrm{DT,0} \left(\frac{t+\tau_\mathrm{d}}{\tau_\mathrm{d}}\right)^{-2},
	\label{eq:dyn_tide}
\end{equation}
where $\Gamma_{\mathrm{DT,0}}$ is the initial dynamical torque and $\tau_\mathrm{d}=t_\Gamma$ is the time after which the dynamical tides rapidly decay. For solar-likes star such as Kepler-221 we put $\tau_\mathrm{d}=460\,\mathrm{Myr}$ \citep{AhuirEtal2021}.  \Fg{toeque_b} plots these expressions for $\Gamma_\mathrm{ML}$ and $\Gamma_{\mathrm{DT}}$ by the solid green and blue lines, respectively.  The torque due to the dynamical tide on planet b is represented by the blue solid line \citep{AhuirEtal2021}, which is much stronger than the equilibrium tides represented by the black line in \fg{toeque_b} (assuming $Q_\star'=10^6$).

We have conducted simulations with the above time-dependent torque expressions acting on planet b, where we vary $\Gamma_\mathrm{ML,0}$ and $\Gamma_\mathrm{DT,0}$. We find that if the masses of the planets follow \texttt{M1} in \tb{exp_para}, torque expressions that are approximately a factor 2 greater than the reference expressions could obtain the desired configuration where planet d ends up at the observed period ratio. These curves are shown with dashed lines in \fg{toeque_b}. From the figure, it is clear that the torque required in the simulation is of the same order of magnitude as the torque generated by mass loss or dynamical tides. After 500 Myr, the increase in $P_\mathrm{d}/P_\mathrm{c}$ slows down, allowing the system to execute an S-curve bend as illustrated in \fg{orb_exp_suc} d. Finally, the planet period ratios increase to the observed values while avoiding the $(3,5,8)$ 3BR line.

\subsubsection{Mass constraint on b, c, and e}
\label{sec:allow_mass}

A more direct way to avoid crossing the \cde{} (3,5,8) 3BR is to assume different masses of the planets. The masses of planets b, c, and e determine the slope of expansion of $P_\mathrm{d}/P_\mathrm{c}$ and $P_\mathrm{e}/P_\mathrm{d}$ defined as $a_\mathrm{cde}$ in \eq{slope_for}. To reach the observed period ratio at the end of the orbital expansion phase, the slope should satisfy $a_\mathrm{cde}>1.56$ (see \se{expansion} and \App{cde-slope}). Therefore, if there are no additional mechanisms (for instance \se{orb_outmig}) acting on the planets, we can constrain the masses and densities of the planets with the current period ratio in the orbital expansion phase.

\begin{figure}
	\centering
	\includegraphics[width=\columnwidth]{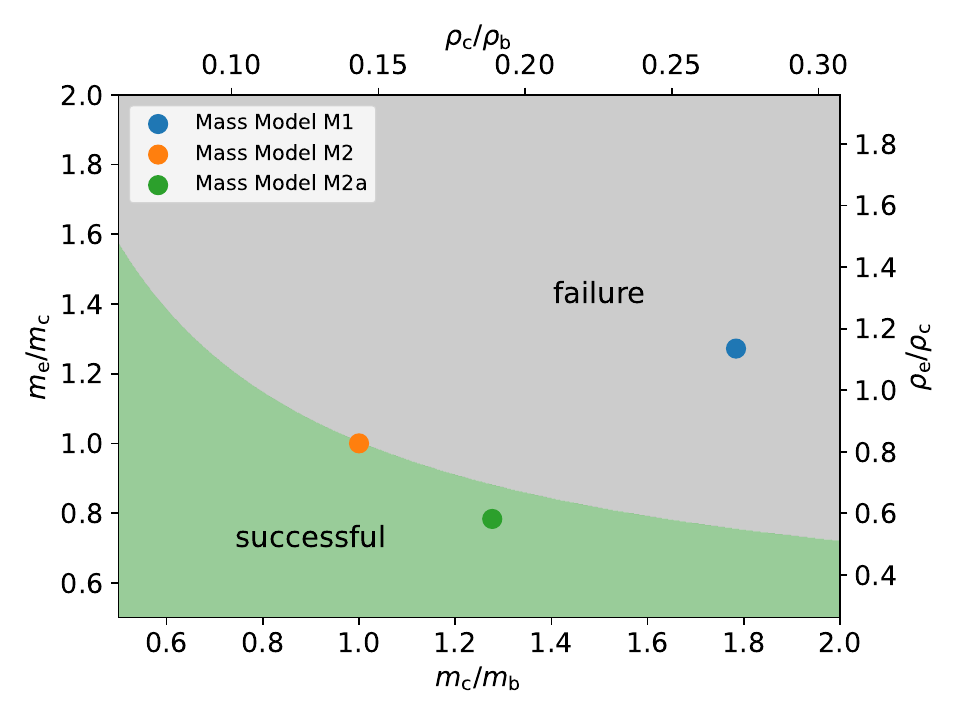}
	\caption{\label{fig:mass_region}Constraints on planet mass and density ratios in the orbital expansion phase with the radius of the planets according to \tb{Kepler-221}. The green region corresponds to the area satisfying $a_\mathrm{cde}>1.56$, where orbital expansion to the observed period ratio is possible, while the planets cannot reach the observed period ratio if the planet mass ratio falls in the gray region. The colored dots represent different mass models in \tb{exp_para}. The conversion to density ratios (upper and right axes) has assumed the radii of the planets according to \tb{Kepler-221}. }
\end{figure}

\Fg{mass_region} shows the constraints on the mass and density ratios in the orbital expansion phase in which the migration of planet d in the orbital expansion phase has been neglected. Successful orbital expansion is only possible if the combination of mass or density ratios falls in the green region, which corresponds to $a_\mathrm{cde}>1.56$. In general, this requires a more massive planet c than planet e (for instance, mass model \texttt{M2a} or \texttt{M2d} in \tb{exp_para}, green dot in \fg{mass_region}). Because mass model \texttt{M2a} and \texttt{M2d} have the same mass for planets b, c, and e, only one dot is presented in \fg{mass_region}. Initial masses according to the mass-radius relationship in \citet{ChenKipping2017} would result in too small $a_\mathrm{cde}$ for successful orbital expansion (\texttt{M1} in \tb{exp_para}, blue dot in \fg{mass_region}) and require additional mechanisms (for instance, stronger damping on planet d and outward torque on planet b, see \se{expansion}).

The peas-in-a-pod mass model (\texttt{M2} in \tb{exp_para}; orange dot in \fg{mass_region}) assumes equal masses for all planets, resulting in $a_\mathrm{cde}=1.56$, just sufficient according to \eq{slope_for2}. Yet, it would be difficult to reach the observed period ratio, because planet d would also migrate slightly outward due to secular interactions, resulting in a deviation from the observed period ratio during expansion (see \App{d-mig}). Therefore, a modest adjustment of the masses -- still within the peas-in-a-pod framework -- is necessary. For example, mass model \texttt{M2a} or \texttt{M2d} with a 25\% more massive planet c would ensure a successful orbital expansion phase.

\section{Discussion}
\label{sec:discussion}

\begin{table*}
\caption{\label{tab:conclusion_tb}Assessment of the dynamical model for the Kepler-221 system. Comments for different scenarios in different phases are listed.}
\centering
\small
\begin{tabular}{l l p{3cm} @{}l@{\;} p{4cm} l p{60mm}}
\hline
Phase & Section & Constraint & & Condition/mechanism & & Comment \\
\hline
I & \ref{sec:phase12}  & First-order resonance chain with five planets  & & Large scale migration & $\bullet$ & Formation of first-order 2BR is a natural result of convergent migration\\
I/II & \ref{sec:phase12} & Planets deep in resonance  & & Needs high eccentricities for planets & $\bullet$ & High eccentricities ensure quick merger after resonances break\\
II & \ref{sec:phase12} & Breaking of resonance chain & & Requires instability in the system & $\bullet$ & Likely happens in many other systems  \\
III & \ref{sec:P3-mec} & 3BR reformation\dotfill & 1. & No effects   & $-$ & Zeroth-order \bce{} 3BR never reforms \\ 
& & & 2. & Anomalous damping on planet c (large $Q_\mathrm{c}/Q_\mathrm{d}$) & $\bullet$ & Can lead to convergent migration but no clear reason why planet c stands out \\ 
& & & 3. & Positive torque on planet b & + & Mechanism can also be applied to Phase IV  \\ 
IV & \ref{sec:orb_parstudy} & Avoid \cde{} 3BR\dotfill & 1. & Positive torque on planet b & $\bullet$ & Mechanism can also be applied to Phase III \\
& & & 2. & Different assumption of the planet masses & + & Needs no additional mechanisms and agrees with peas-in-a-pod scenario \\
IV & \ref{sec:age_con} & Long-term expansion\dotfill & 1. & System age much older & $\bullet$ & No need for mechanisms like obliquity tides but inconsistent with age inference of Kepler-221 \\
& & & 2. & Obliquity tides & $\bullet$ & Can explain the fast orbital expansion but hinders 3BR reformation in Phase III\\
\hline
\end{tabular}
\end{table*}

\subsection{Assessment of the model}
\label{sec:as_model}

In this work, we have investigated several mechanisms to explain the observational constraints that the Kepler-221 system presents to us. A summary of the comments for each of these mechanisms is listed in \tb{conclusion_tb} in chronological order. In Phase I and II, the merging of planets $\mathrm{d}_1$ and $\mathrm{d}_2$ is a natural result of the first-order resonance breaking after disk dispersal, and a quick merging within hundreds of years is achieved with high eccentricities for planets after disk dispersal (see \se{phase12}). 

Perhaps the most critical phase for the collision scenario is the re-establishment of the 6:3:1 nonadjacent 3BR between planets b, c, and e. We found that reformation is not spontaneous \citep{Petit2021}, but instead requires either a stronger damping on planet c or a positive torque on planet b. The latter scenario is favored because the outward migration of planet b also promotes a successful subsequent orbital expansion phase, which ensures that planets c, d, and e can avoid 3BR encounters during Phase IV (see \se{orb_outmig}). Still, resonance reformation is not guaranteed, but stochastic, and successful only in ${\sim}17\%$ of our simulations (see \se{P3-mec}).

It is likely that planets b, c, and e have experienced significant orbital expansion while locked in the 3BR configuration. If the system is young, as suggested by \citet{BergerEtal2018}, eccentricity tides alone are unlikely to be the sole driver responsible for the 3BR expansion. Therefore, additional damping mechanisms, such as obliquity tides \citep{GoldbergBatygin2021}, are required to accelerate the expansion. However, a too-strong dissipation would hinder the reformation of \bce{} 3BR in Phase III. This tension between Phase III (which favors weak damping) and Phase IV (which favors robust damping) is discussed below in \se{age_con}.

Also, we found in this work that planet d plays an instrumental role in this expansion as it has the potential to kick the planets out of resonance (see \se{orb_exp_mec}). This outcome could be avoided by tuning the parameters for the outward migration of planet b, but a more straightforward explanation is that prolonged expansion was made possible by the planet's mass ratios.

\subsection{Mass constraints}
\label{sec:mass_con}

Our modeling provides constraints on the yet undetermined masses of the Kepler-221 planets \citep{BergerEtal2018}. The first clue comes from the formation scenario (Phase II-III), where it was postulated that planet d arises from the merger of two planets. The merging could have attributed to significant atmosphere loss but little mass loss \citep{GhoshEtal2024, DouEtal2024}. If so, it is expected that the density of planet d is higher, which would make planet d the most massive planet in the current system. The reformation of 3BR in the post-collision phase is still viable with planet d twice as massive as other planets (see \se{suc_rate}). Furthermore, once the \bce{} resonance is established, the long-term evolution of the system (Phase IV) is mostly independent of planet d's mass (see \App{d-mig}). Therefore, a more massive planet d remains a possibility.

The second stronger clue comes from the orbital expansion phase. This clue is stronger as it is independent of the formation model. In the orbital expansion phase, a slope of $a_{\mathrm{cde}}>1.56$ in the \cde{} orbital plane (see \se{orb_exp_mec}) is required, which constrains the masses and densities of planets b, c, and e in the absence of additional dissipative forces or torques on planets (see \se{allow_mass}). From this, similar masses with a slightly more massive planet c are expected, such as in model M2a, consistent with the peas-in-a-pod model \citep{WeissEtal2020, WeissEtal2023}. This model renders the densities of the planets progressively decrease with distance, $\rho_{b}>\rho_{c}>\rho_{e}$ (see \fg{mass_region}). A possible origin of the similar masses and the decreasing density of the planets in the system could be due to photoevaporation. Inflated low-mass planets with orbits of a few tenths of an AU could experience a boil-off phase after disk dispersal, in which the planets lose their envelope and contract quickly with only a few of their original envelope left after the disk dispersal \citep{OwenWu2016}. Afterwards, an X-ray and EUV-dominated photoevaporation could further erode the planet atmosphere with a timescale of approximately $100\,\mathrm{Myr}$ \citep{OwenWu2013, LopezFortney2013}. One example of systems with photoevaporation-dominated planets is Kepler-36, which hosts two close-in planets (closer than $0.15\mathrm{au}$) with similar atmosphere fractions when they are born and the inner planet has lost almost all of its atmosphere \citep{CarterEtal2012, LopezFortney2013, Owen2019}. A similar photoevaporation process could have happened in the Kepler-221 system for planet b, resulting in a planet b of terrestrial density and puffier outer planets. An additional advantage of photoevaporation on planet b is that the associated rapid mass loss would contribute to the outward migration of planets b and c, which facilitates success in Phase III and Phase IV of our model (see \ses{orb_outmig}{P3-mec}).

\begin{figure}
\centering
\includegraphics[width=\columnwidth]{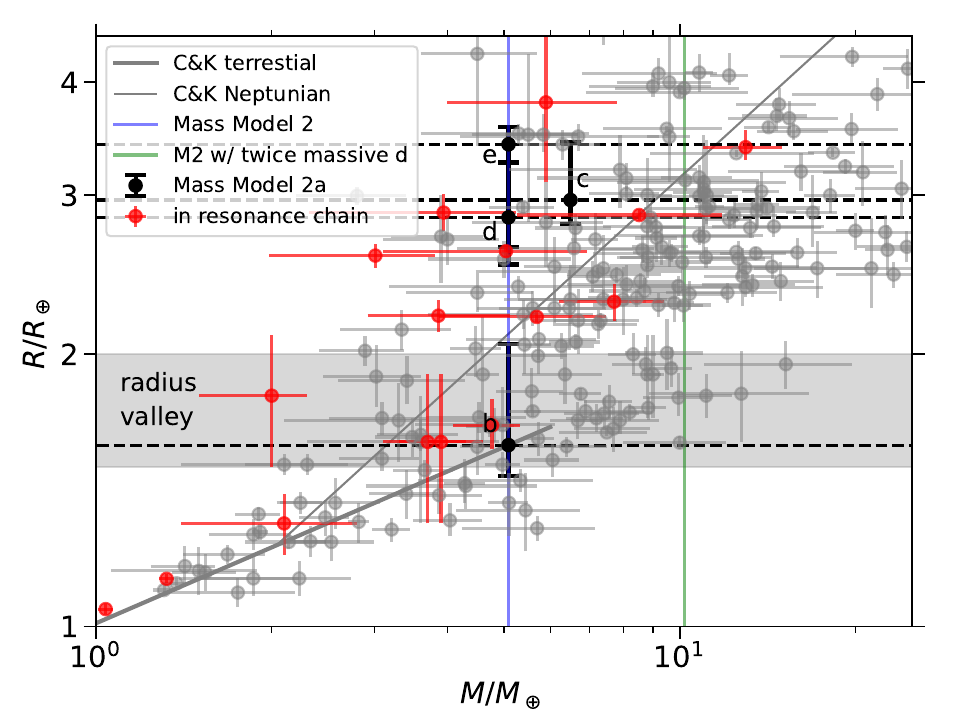}
\caption{Mass-radius diagram of the Kepler-221 planets with mass models listed in \tb{exp_para} along with the exoplanets sample. Black dots and error bars represent the masses of planets according to mass model M2a of \tb{exp_para} and radii according to \tb{Kepler-221}. The blue line shows the masses of the planets according to mass model M2 of \tb{exp_para}. The gray lines provide the mass-radius relation according to \citet{ChenKipping2017}. The green line indicates the mass of d if it would be twice as massive as the other planets, following mass model M2d of \tb{exp_para}. The background dots and error bars show the masses and radii of other exoplanets, with those in red representing planets in a first-order resonance chain involving at least three planets. The gray area shows the radius valley between $1.5R_\oplus$ and $2.0R_\oplus$ according to \citet{FultonEtal2017}.}
\label{fig:MR_compare}
\end{figure}

Assuming model M2a of \tb{exp_para}, \fg{MR_compare} plots the planets in the Kepler-221 system along with the planets of the exoplanet sample for which masses and radius are available.\footnote{https://exoplanetarchive.ipac.caltech.edu} The plot shows that assuming similar masses following the peas-in-a-pod model \citep{WeissEtal2020, WeissEtal2023} in M2a in \tb{exp_para} results in planet c and e puffier than Neptune. However, such an assumption is not unreasonable because similar super-puff planets exist in the exoplanet sample, especially in resonance chains (red background dots in \fg{MR_compare}). For example, TOI-178 hosts six planets in a first-order resonance except for the innermost planets \citep{LeleuEtal2021TOI}. Among these planets, TOI-178 d and g are super-puff planets with radii similar to and density lower than the planets c and e estimated in M2a in \tb{exp_para} \citep{LeleuEtal2021TOI, DelrezEtal2023}. Kepler-51 hosts three planets in a 3:2:1 resonance chain, and two of the planets are super-puff planets with masses similar to and radii much larger than the Kepler-221 estimation for planets c and e in mass models M2a. Therefore, the outer planets in the Kepler-221 system could potentially be another sample of super-puff planets in resonance. 

It is worthwhile to conduct RV measurements on the Kepler-221 system with facilities such as Keck \citep{VogtEtal1994}. The significance of the RV detection is as follows: (i) if planet d turns out to be more massive than the other planets, it suggests a collisional origin. (ii) If the mass ratios of the planets b, c, and e fall in the green region in \fg{mass_region}, it offers the most natural scenario to explain the sustained orbital expansion phase (see \se{expansion}). And (iii) it can provide evidence that planets c and e belong to the super-puff planets, while planet b is on the other side of the radius valley \citep{FultonEtal2017} (see \fg{MR_compare}).

\subsection{Age constraint}
\label{sec:age_con}

Kepler-221 is a young system with an age below 650 Myr inferred from its large lithium abundance \citep{BergerEtal2018, GoldbergBatygin2021}. 
Therefore, planets b, c, and e in 3BR need to undergo a sufficiently rapid orbital expansion to reach their present period ratios.

We define the time required to expand the b, c, and e 3BR from the integer period ratio to the observed period ratio as $t_\mathrm{exp}$. In our simulation, $t_\mathrm{exp}$ mainly depends on the tidal damping strength on the planets (determined by $Q_\mathrm{phy}$) and is hardly affected by other parameters such as the mass ratios of the planets. For example, panels e and f in \fg{orb_exp_suc} show a successful expansion with masses of the planets according to mass model M2a in \tb{exp_para} and the same $Q_\mathrm{phy}$ for all planets. From the orbital expansion phase simulations (see \se{expansion}), the relation between $t_\mathrm{exp}$ and $Q_\mathrm{phy}$ can be expressed as:
\begin{equation}
    t_\mathrm{exp}\approx 380Q_\mathrm{phy} \,\mathrm{Myr},
\label{eq:exp_time}
\end{equation}
which follows from the linearity of the damping rate with $Q$. Because the expansion duration time $t_\mathrm{exp}$ should not exceed the age of the system, we can conclude $Q_\mathrm{phy}<2$ corresponding to the maximum age of $650\;\mathrm{Myr}$ for the Kepler-221 system. 

Such results conflict with the formation model of Kepler-221 in two aspects. Firstly, The value of $Q_\mathrm{phy}$ is over two orders of magnitudes smaller than the typical value of the tidal quality factor $Q$ for terrestrial planets \citep{LeeEtal2013, SilburtRein2015}, indicating the existence of other much more efficient dissipation mechanisms to speed up the simulation. Secondly, the reformation process of 3BR requires gentle convergent migration between planets \bce{} corresponding to relatively large $Q_\mathrm{phy}$ (in Phase III we apply $Q_\mathrm{phy}>3$, see \tb{fig_para} and \se{suc_rate}). This conflicts with the $Q_\mathrm{phy}<2$ constraint derived from the age of the system.

\begin{table*}
\centering
\caption{\label{tab:K2-138}Dynamical configuration of K2-138 planetary system according to \citet{ChristiansenEtal2018} and \citet{LopezEtal2019}. }
\renewcommand{\arraystretch}{1.2}
\small
\begin{tabular}{ccccccc}
\hline
Planet                & b & c & d & e & f & g \\ \hline
Radius($R_\oplus$)             & $1.510^{+0.110}_{-0.084}$  &  $2.299^{+0.120}_{-0.087}$ &  
	$2.390^{+0.104}_{-0.084}$ & 
	$3.390^{+0.156}_{-0.110}$  &  
	$2.904^{+0.164}_{-0.111}$ & 
	$3.013^{+0.303}_{-0.251}$  \\
Mass($M_\oplus$)               & $3.10\pm1.05$   & 
	$6.31^{+1.13}_{-1.23}$  & $7.92^{+1.39}_{-1.35}$  &  $12.97^{+1.98}_{-1.99}$ & $1.63^{+2.12}_{-1.18}$  & $4.32^{+5.26}_{-3.03}$  \\
Period(days)          & $2.353$  & 
	$3.560$  & $5.405$   &  
	$8.261$ & 
	$12.758$  &  
	$41.968$\\
Period Ratio to Inner Planet & \textbackslash{}  &  1.513 & 1.518  &  1.529 & 1.544  &  3.290 \\ 
$(p,q,r)$ of 3BR\tablefootmark{a}    &   &   (2,3,5) &   (2,3,5) &   (2,3,5) &   (2,3,4)\\
Normalized B-values of 3BR\tablefootmark{b} & & $1.77\times10^{-3}$ & $9.77\times10^{-4}$ & $2.04\times10^{-4}$ & $4.60\times10^{-4}$ \\ \hline
\end{tabular}
\end{table*}

\begin{table*}
\centering
\small
\caption{\label{tab:Kepler-402}Dynamical configuration of the Kepler-402 planetary system according to \citet{RoweEtal2014}. }

\begin{tabular}{ccccc}
\hline
Planet                       & b & c & d & e \\ \hline
Radius($R_\oplus$)                    &  
	$1.22\pm0.24$ & $1.56\pm0.35$   &  $1.38\pm0.27$  &  $
	1.46\pm0.29$ \\
Period(days)                 & 4.029  & 6.125  &  8.921 &  11.243 \\
Period Ratio to Inner Planet & \textbackslash{}  & 1.520  & 1.457  & 1.260  \\ 
$(p,q,r)$ of 3BR\tablefootmark{a}    &   &   (3,5,8) &   (5,11,16) \\
Normalized B-values of 3BR\tablefootmark{b} & & $6.09\times10^{-3}$ & $1.05\times10^{-2}$ \\ \hline
\end{tabular}

\tablefoot{ 
\tablefoottext{a}{Same definition as \tb{K2-138}.} \\
\tablefoottext{b}{The non-adjacent planets b, c, and e are close to $(7,8,15)$ 3BR with a small normalized B-value of $9.25\times10^{-7}$.}
}
\end{table*}

\citet{GoldbergBatygin2021} point out that the timescale puzzle could be solved with some additional dissipation mechanism required (for instance, obliquity tides; \citealt{MillhollLaughlin2019}). Obliquity tides occur when planets have a large axial tilt
(obliquity) \citep{MillhollLaughlin2019} and have been invoked to speed up the orbital expansion in the Kepler-221 system \citep{GoldbergBatygin2021}. A non-zero obliquity for planets can be maintained if the planet is in Cassini state where it reaches a secular spin-orbit resonance with synchronized precession of planetary spin and orbital angular momentum. The total energy dissipation rate for a planet with non-zero eccentricity and obliquity can be expressed as \citep{MillhollLaughlin2019}:
\begin{equation}
    \dot{E}(e,\epsilon)=\frac{2K}{1+\cos^2\epsilon}\left[\sin^2\epsilon+e^2\left(7+16\sin^2\epsilon\right)\right],
\label{eq:ob_tide}
\end{equation}
where $e$ is the eccentricity of the planet and $\epsilon$ is the obliquity of the planets. Here, $7Ke^2$ is the tidal dissipation in the absence of obliquity ($\epsilon=0$), where $K$ depends on the usual stellar and planet properties \citep{MillhollLaughlin2019}. 

Planets are likely to have low obliquity during the disk phase as their angular momentum and spin vectors are aligned with the disk. But during the brief phase of dynamical instability and the resonance reformation afterward, planet b or c could have acquired a mutual inclination with respect to the other planets, which is a requirement for trapping in a Cassini state. Additionally, planet c dominates the tidal expansion process at the same level as planet b due to the large radius of c compared to b (see \eq{te_exp}). \Eq{ob_tide} demonstrates that even modest obliquity (for example, on planet c, consistent with \se{P3-mec}) can dominate the energy dissipation, and drive the tidal expansion at rates much higher than what can be obtained from tidal damping alone.

However, the drawback of a small $Q_\mathrm{phy}$ is that such strong tidal damping would also lead to a rapid migration during the 3BR reformation phase. This makes the 3BR reformation unlikely (see \se{suc_rate}). This implies either that the age of the system is actually older, or that some unknown mechanisms have sped up the orbital expansion in Phase IV. For example, \fg{all_8} is a successful simulation throughout all phases, which corresponds to a system age of around $2\,\mathrm{Gyr}$. On the other hand, the orbital expansion configuration in Phase IV is unrelated to the expansion speed. Therefore, the constraints on the masses of the planets are solid regardless of the system age (see \se{allow_mass}).

\subsection{Application to other systems}
\label{sec:other_system}

Other exo-planetary systems with one planet outside the first-order resonance chain may also be described with ideas introduced in this work. In particular, we highlight the K2-138 and Kepler-402 systems with the information of these systems listed in \tb{K2-138} and \tb{Kepler-402}. K2-138 is a K1-type star with six super-Earths. The period ratios between the inner five planets are close to 3:2 and all planets in K2-138 are in three-body resonance, with the innermost three pairs in $(p,q,r)=(2,3,5)$ 3BR. The outermost three planets (planets e, f, and g) are locked in a peculiar first-order $(2,3,4)$ 3BR, which is the first pure first-order 3BR detected in a multi-planetary system \citep{CerioniBeauge2023}. Kepler-402 is an F2-type star with four super-Earths (planets b, c, d, and e). Similar to Kepler-221, planets b, c, and e are in a $(7,8,15)$ 3BR, with the inner period ratio close to 3:2 and the outer period ratio close to 16:9. Planet d is not in resonance with any other planet.

The 3BRs in the K2-138 system are listed in \tb{K2-138}. The inner three 3BRs pairs are close to $(2,3,5)$ 3BR, with the period ratio increasing from inner pairs to outer pairs. Therefore, it is reasonable to assume that the inner five planets are originally locked into 3:2 two-body resonances with each adjacent triple also locked in 3BRs. At 1.544 and 3.290 the period ratios of planets e, f, and g are relatively far from integer period ratios, which could be the result of orbital expansion along $(2,3,4)$ first-order 3BR from an inner 3:2 and outer 3:1 two-body resonance. The formation of first-order 3BRs was also seen in our simulations for the Kepler-221 system when we attempted to reform the \bce{} 3BR in the post-collision phase (see \se{collision} and \App{1st3BR}). This demonstrates that the formation of pure first-order 3BR is possible in systems similar to Kepler-221. Compared to zeroth-order 3BRs, a first-order 3BR does not originate from two two-body resonances and could form relatively far from exact 2-body commensurabilities \citep{Petit2021} (see \App{1st3BR}). A possible formation scenario of the K2-138 could be that the inner five planets are first locked in a chain of 3:2 two-body resonances. Then tidal damping expands the period ratio of the inner planet, and the outer planet g coincidentally comes close to the $(2,3,4)$ first-order 3BR and becomes trapped, similar to the resonance reformation process discussed in \se{P3-mec} and \App{1st3BR}.

The period ratios of the Kepler-402 planetary system are shown in \tb{Kepler-402} with the corresponding closest 3BR. It is clear from the figure that the adjacent planets are far from 3BR with larger normalized B-value but planets b, c, and e are in the $(7,8,15)$ pure 3BR relatively far from integer period ratio, similar to the Kepler-221 system. This 3BR between planets b, c, and e might expand from an inner 3:2 and outer 16:9 two-body resonance. However, it is unlikely that such a high-order resonance can be formed with simple convergent migration in the disk phase. Also, the middle planet d makes the resonance formation between c and e harder (see \se{collision}). Therefore, the $(7,8,15)$ 3BR for planets b, c, and e probably does not originate from two-body resonance between c and e. Possibly, the Kepler-402 system formed similarly to Kepler-221 in the sense that that there were originally two planets between planets c and e ($\mathrm{d}_1$ and $\mathrm{d}_2$) with all five planets in a first-order resonance chain. The resonance number for planets $\mathrm{d}_1$ and $\mathrm{d}_2$ two-body resonance is very high (7:6 or 8:7 resonance, facilitating the resonances to break quickly after disk dispersal \citep{MatsumotoEtal2012}, and resulting in the merging of planets $\mathrm{d}_1$ and $\mathrm{d}_2$ into planet d. Afterward, tidal damping reforms the 3BR between planets b, c, and e, and the system period ratios expand to the observed value.

\section{Conclusion}
\label{sec:conclusion}

We have proposed a multi-phase formation model for the dynamical history of the Kepler-221 system, in line with its present architecture, of which the most peculiar feature is the out-of-resonance intermediate planet d. The envisioned scenario relies on two Ansatzes. First, there were originally five planets (planets b, c, $\mathrm{d}_1$, $\mathrm{d}_2$, and e) in a chain of first-order resonances. Second, after disk dispersal, the system experienced an instability, which broke all resonances and caused planets $\mathrm{d}_1$ and $\mathrm{d}_2$ to merge into planet d with its position consistent with its observed location. Under dissipative processes such as tidal damping, the 6:3:1 three-body resonance between planets b, c, and e reformed. On evolutionary timescales, the period ratio of these three planets expanded under tidal dissipation until they reached the observed period ratio.

To investigate the feasibility of the model we have carried out a detailed parameter study, in a modular fashion, where we detail the conditions necessary to overcome each milestone. The conclusions that emerge from this study are the following: 

\begin{enumerate}
	\item Immediately after the collision of d$_1$ and d$_2$, convergent migration between planets b and c is essential to reform the \bce{} 3BR. This implies a positive torque operating on planet b or an anomalously small tidal-Q of planet c. Stronger damping on planet c (possibly due to strong obliquity tides \citep{GoldbergBatygin2021}) would migrate planet c inward while a positive torque on planet b would migrate planet b outward. If the appropriate conditions are satisfied, the success rate amounts to 17\% (see \fg{tree_plot}).
	\item If these conditions are not present, resonance reformation would fail or the \bce{} planets would end up in a $(1,3,3)$ first-order three-body resonance.
	\item The properties of planet d are crucial for the outcome of the orbital expansion of the \bce{} sub-system. In order for the planets' period ratios to evolve toward their current values, planets c, d, and e must have avoided the $(3,5,8)$ resonance, which would break the b, c, and e 3BR with a probability of 95\%.
	\item Therefore, planets \bce{} should have evolved along a line of slope $a_\mathrm{cde}>1.56$ in the period ratio plane of $P_\mathrm{e}/P_\mathrm{d}$ vs $P_\mathrm{d}/P_\mathrm{c}$. Conservation of angular momentum and the \bce{} 3BR angle gives an analytical expression for $a_\mathrm{cde}$ (see \eq{slope_for}), which only depends on the mass ratios of planets b, c, and e.
	\item The condition $a_\mathrm{cde}>1.56$ constrains the mass ratios of planets b, c, and e in the Kepler-221 system (see \se{mass_con}) to values consistent with the peas-in-a-pod scenario. Satisfying this mass constraint, a sustained expansion of the system is guaranteed. In addition, a positive torque on planet b could have increased $a_\mathrm{cde}$ to ensure successful orbital expansion when the mass constraint is not satisfied. This mechanism is particularly attractive as it also promotes resonance reformation.
	\item Resonance reformation requires the planet to gentle converge onto each other at rates that correspond to effective damping parameter $Q_\mathrm{phy}>3$. On the other hand, the presumably young age of the system would require effective $Q_\mathrm{phy}<2$. This could imply that either the system actually is older, or that some unknown mechanisms have sped up the orbital expansion process in Phase IV.
	\item The dynamical model proposed for the Kepler-221 system can be applied to other planetary systems in resonance, in particular, K2-138, which features three planets in a pure first-order 3BR, and Kepler-402, which, as with Kepler-221, features an intermediate planet not part of the resonance chain.
\end{enumerate}

\section*{Data availability}
The data underlying this article will be shared on reasonable requests to the corresponding author.

\begin{acknowledgements}
    We thank the referee for their constructive comments, which significantly improved the presentation of this work. This work is supported by the National Natural Science Foundation of China under grant No. 12250610189, 12233004 and 12473065. The authors gratefully thank Sharon Xuesong Wang, Fei Dai, Haochang Jiang, Yu Wang and Helong Huang for their useful discussions. Software: matplotlib \citep{Hunter2007, CaswellEtal2021}, REBOUND \citep{ReinLiu2012}, REBOUNDx \citep{TamayoEtal2020}.
\end{acknowledgements}



\bibliographystyle{aa}
\bibliography{ads} 




\appendix

\onecolumn

\section{Trapping in first-order 3BR}
\label{app:1st3BR}

In the post-collision phase, planets b, c, and e can escape from zeroth-order 3BR and instead get trapped in first-order 3BR, as shown in panels a and d in \fg{123_mech}. Although both originate from b/c 2:1 and c/e 3:1 2BR, the definitions of zeroth-order and first-order 3BR for planets b, c, and e are different. The expression for b/c 2:1 outer and c/e 3:1 inner 2BR angle are:
\begin{align}
	 & \phi_\mathrm{bc,o}=\lambda_\mathrm{b}-2\lambda_\mathrm{c}+\varpi_\mathrm{c}  \\
	 & \phi_\mathrm{ce,i}=\lambda_\mathrm{c}-3\lambda_\mathrm{e}+2\varpi_\mathrm{c},
	\label{eq:bce_2BR}
\end{align}
where $\lambda_\mathrm{i}$ is for mean longitude for the planet $i$ and $\varpi_\mathrm{c}$ represents the longitude of periapsis of planet c. For the definition of zeroth-order 3BR, $\varpi_c$ is eliminated:
\begin{equation}
	\phi_\mathrm{bce,0}=2\phi_\mathrm{bc}-\phi_\mathrm{ce}=2\lambda_\mathrm{b}-5\lambda_\mathrm{c}+3\lambda_\mathrm{e},
	\label{eq:app_0th_3BR}
\end{equation}
The definition of a first-order 3BR contains the argument of periapsis of one planet and requires that $p+q-r=1$ (see \eq{3BR_define}). One example of first-order 3BR for planets b, c, and e is:
\begin{equation}
	\phi_\mathrm{bce,1}=\phi_\mathrm{bc}-\phi_\mathrm{ce}=\lambda_\mathrm{b}-3\lambda_\mathrm{c}+3\lambda_\mathrm{e}-\varpi_\mathrm{c},
	\label{eq:app_1st_3BR}
\end{equation}

If planet b/c and c/e are in a 2BR resonance (i.e., $\phi_\mathrm{bc,o}$ and $\phi_\mathrm{ce,i}$ librate), both $\phi_\mathrm{bce,0}$ and $\phi_\mathrm{bce,1}$ librate because these two 3BR angles are linear combination of the 2BR angle. In this case, planets b, c, and e are trapped in $(2,3,5)$ zeroth-order 3BR (see \eq{3BR_define}). If $\phi_\mathrm{bce,1}$ liberates, but not the zeroth-order 3BR angle and 2BR angles, then planets b, c, and e are trapped in $(1,3,3)$ first-order 3BR. The strength of 0th and first-order 3BRs are roughly determined by $p+q$, with the smaller value representing stronger resonance \citep{Petit2021}.

\begin{figure*}[h!]
	\sidecaption
	\includegraphics[width=0.7\textwidth]{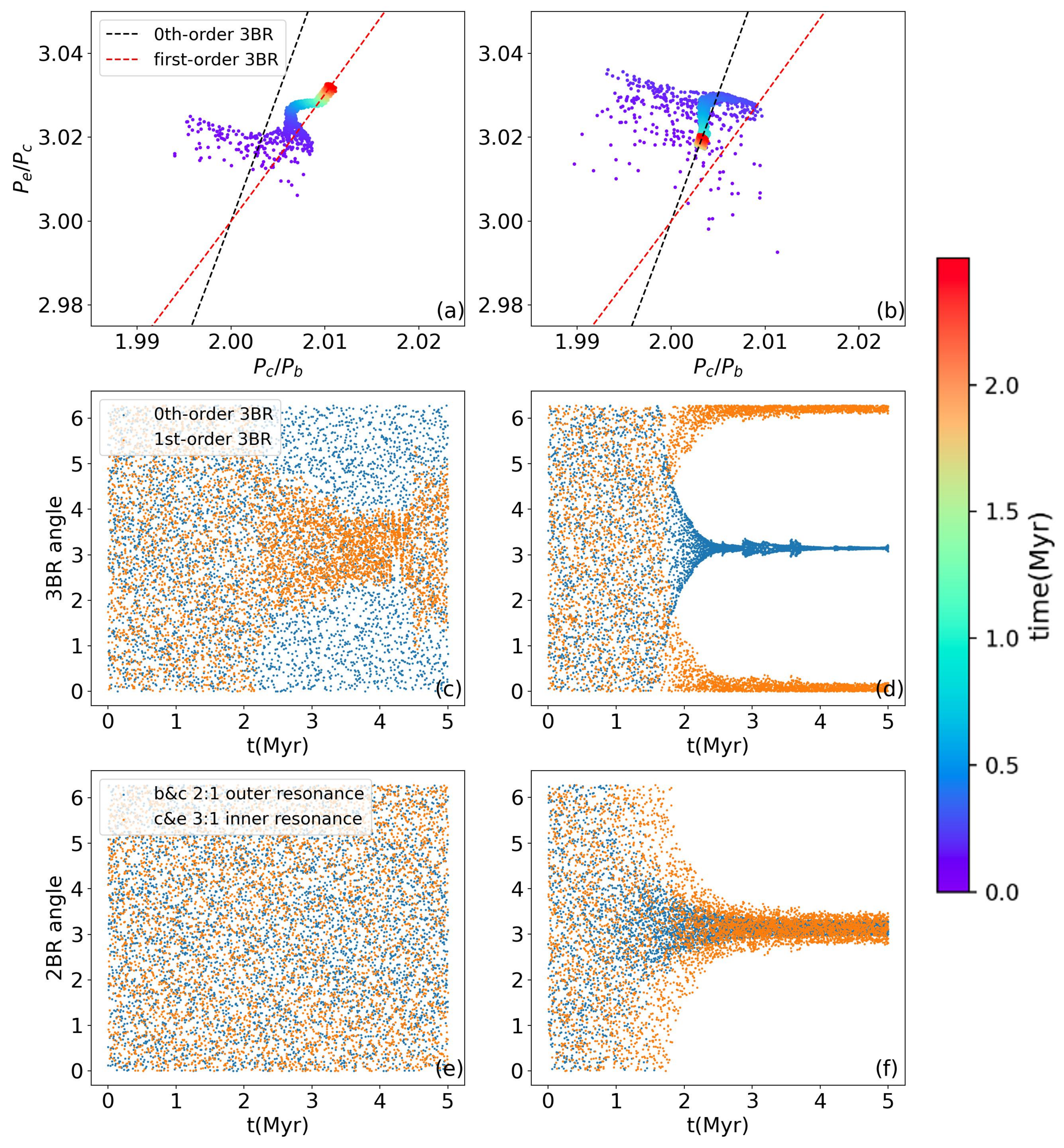}
	\caption{\label{fig:1st_3BR_fig} Period ratios and resonance angles for planets b, c, and e trapped in 0th and first-order 3BR in post-collision phase. Panels a and b represent the period ratio evolution with the color bar indicating time. The black and red dashed lines respectively correspond to $(2,3,5)$ zeroth-order 3BR and $(1,3,3)$ first-order 3BR following \eq{3BR_ratio}. Panel c and d plot the evolution of the zeroth-order 3BR angle $\phi_\mathrm{bce,0}=2\lambda_\mathrm{b}-5\lambda_\mathrm{c}+3\lambda_\mathrm{e}$ and first-order 3BR angle $\phi_\mathrm{bce,1}=\lambda_\mathrm{b}-3\lambda_\mathrm{c}+3\lambda_\mathrm{e}-\varpi_\mathrm{c}$. Panel e and f plot the b/c outer resonance angle $\phi_\mathrm{bc,o}=\lambda_\mathrm{b}-2\lambda_\mathrm{c}+\varpi_\mathrm{c}$ and c/e inner resonance angle $\phi_\mathrm{ce,i}=\lambda_\mathrm{c}-3\lambda_\mathrm{e}+2\varpi_\mathrm{c}$. The right panels correspond to the simulation where a positive torque is applied on planet b and planets b, c, and e get trapped in zeroth-order 3BR. In contrast, in the left panels, there is no torque on planet b and planets are trapped in a first-order 3BR.}

\end{figure*}

Generally, the formation of zeroth-order 3BR is an inevitable result of 2BR formation, so planets should first be captured when their period ratios are close to integer ratios. If planets are far away from integer ratio, then they are more likely to be captured into the stronger first-order 3BR with smaller $p+q$ values for planets b, c, and e. \Fg{1st_3BR_fig} illustrates the process of resonance formation in the post-collision phase (see \se{collision}). The left panels show the simulation where no additional torque is applied on planet b, and the tidal $Q$ parameter is the same for all planets. Planet b, c, and e skip the zeroth-order 3BR and get trapped into first-order 3BR. This is because planets b and c did not first reform 2BR because they do not undergo convergent migration \citep{Petit2021}. So the planets period ratio expands skipping the zeroth-order resonance (black dashed line in panel a), and forms the stronger first-order 3BR along the red dashed line. Therefore, only the first-order resonance angle librates (see panels c and d in \fg{1st_3BR_fig}).

The right panel in \fg{1st_3BR_fig} corresponds to a simulation with an additional positive torque on planet b (see \se{P3-mec}), which results in the outward migration of planet b. This decreases the period ratio between planets b and c (panel b) and leads to convergent migration between these two planets, and first reform b/c 2BR near integer ratio around 1.5\;Myr (panel f). This promotes the reformation of the zeroth-order 3BR, which forms simultaneously with the c/e 2BR (panel d and f) around 1.7\;Myr.

The simulations show that the formation of pure first-order 3BR is possible if planets are close to strong first-order 3BR position due to resonance breaking event. Similar scenarios could be applied to the K2-138 system, which hosts the only pure first-order 3BR resonance identified in the exoplanet census \citep{CerioniBeauge2023}. In the K2-138 system, the outermost three planets are trapped in a pure first-order 3BR. Possibly, the formation of K2-138 first-order 3BR is similar to the resonance formation process in Kepler-221, as shown in \fg{1st_3BR_fig} (see \se{other_system}).

\section{Optimized collision phase}
\label{app:opt_model}

In order to obtain the proper initial conditions for Phase IV in \se{expansion}, we optimized Phase II and III, in which planets $\mathrm{d}_1$ and $\mathrm{d}_2$ are removed and planet d is added back adiabatically. Although artificial, this step ensures that planets b, c, and e remain in resonance and allow us to focus on the ensuing orbital expansion phase. We integrate the simulations of this orbital expansion phase until the point where the resonance breaks and the expansion process is the same as what we observe applying the initial condition directly from Phase III (see \fg{all_8}), which proves the viability of this optimized model. 

The optimized Phase II and III are advantageous to preserve the 3BR while merging planets $\mathrm{d}_1$ and $\mathrm{d}_2$. In this way, we arrive at a compact version of Kepler-221 with planets b, c, and e in 3BR, and it can now be examined whether this configuration can evolve into the present configuration of Kepler-221.

\begin{figure*}[h!]
	\sidecaption
	\centering
	\includegraphics[width=0.7\textwidth]{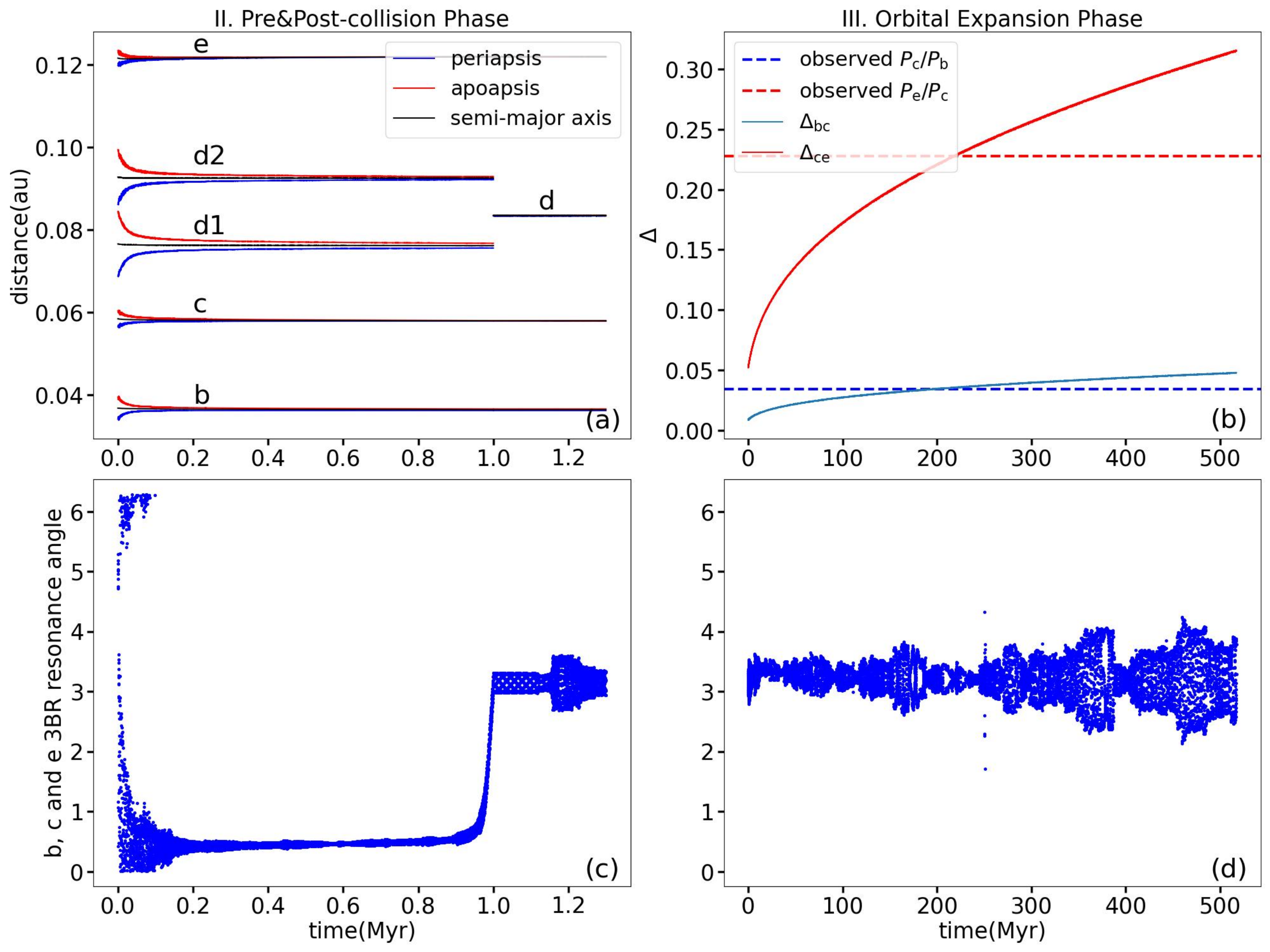}
	\caption{\label{fig:orb_example}Resonance reformation and orbital expansion under optimized conditions with mass model M1 in \tb{exp_para}. Panel a shows the evolution of planets' semi-major axis (black), pericenter (blue), and apocenter (red) in the optimized collision phase, in which planets $\mathrm{d}_1$ and $\mathrm{d}_2$ are adiabatically replaced by planet d. Panel b represents the expansion of period ratio from integer ratio in Phase III, with $\Delta$ defined as $\Delta_\mathrm{bc}=P_\mathrm{c}/P_\mathrm{b}-2$ and $\Delta_\mathrm{ce}=P_\mathrm{e}/P_\mathrm{c}-3$. In panel c and d, the blue dots show the 3BR angle between b, c, and e: $\phi_{3BR}=2\lambda_b-5\lambda_c+3\lambda_e$. The figure shows that the b, c, and e resonance reforms after the simplified collision phase (panels a and c), whereafter the period ratio expands with tidal damping (panels b) before the resonance finally breaks at a larger planet period ratio, ceasing the expansion. }
	\label{fig:orb_exap}
\end{figure*}

Following the model introduced in \se{model}, we proceed with simulation in chronological order with the optimized model, starting from the collision phase. Panels a and c in \fg{orb_example} show an example simulation of the optimized collision phase. We initialize the five planets in first-order resonance with masses according to Model 1 in \tb{exp_para} \citep{ChenKipping2017}. In panel a, the \bce{} 3BR angle is first on a fixed value different from 0 or $\pi$ because multiple 3BR are present for the five planets in the system.  In the optimized collision phase, planets $\mathrm{d}_1$ and $\mathrm{d}_2$ are removed by decreasing their masses adiabatically with a timescale of 1 Myr until their masses reach zero. When $\mathrm{d}_1$ and $\mathrm{d}_2$ are removed, only \bce{} 3BR is left and it goes to the isolated fixed point librating around $\pi$ \citep{Delisle2017}. After this, planet d is inserted at its current position (panel a) and its mass adiabatically increases until $m_\mathrm{d}$ reaches the value shown in \tb{exp_para} with a timescale of 0.3 Myr, which ensures that the 3BR between planets b, c, and e do not break due to the mass change of planet d.

\twocolumn

\section{Mass dependence in Phase IV}
\label{app:cde-slope}
In the orbital expansion phase, the slope of expansion of $P_\mathrm{d}/P_\mathrm{c}$ and $P_\mathrm{e}/P_\mathrm{d}$ is crucial because it determines whether planet d can reach the observed position, as shown in panel b of \fg{orb_exp_fail}. In this section, we derive the analytical expression of the slope (\eq{slope_for}) under the assumption that planet d is stationary.

We first consider the simplified model for the orbital expansion phase, as shown in \fg{appA}. There are two constraints for the starting period ratio in \fg{appA}. Firstly, the starting period ratio of $P_\mathrm{d}/P_\mathrm{c}$ and $P_\mathrm{e}/P_\mathrm{d}$ should be exterior to (to the right of) the $(3,5,8)$ 3BR between planets c, d, and e (the blue dashed line); otherwise the 3BR between planets b, c, and e would break due to the encounter of the c, d, and e 3BR (as explained in \se{expansion}). Secondly, the initial period ratio of c and e should satisfy $P_\mathrm{e}/P_\mathrm{c}=3$ (right of the red dashed line in \fg{appA}). Based on these two constraints, the lower limit of the slope of $P_\mathrm{d}/P_\mathrm{c}$ and $P_\mathrm{e}/P_\mathrm{d}$ for successful expansion to observed period ratio corresponds to the black solid line in \fg{appA}, which starts at the intersection of the two constraint lines (black dot in \fg{appA}) and ends at the observed period ratio, with a slope of 1.56. If we assume planet d does not migrate during the orbital expansion ($P_\mathrm{d}$ is a constant), then the slope can be expressed as:
\begin{equation}
	a_{\mathrm{cde}}=\frac{\Delta({P_\mathrm{e}/P_\mathrm{d}})}{\Delta(P_\mathrm{d}/P_\mathrm{c})}=-\frac{P_\mathrm{c}^2}{P_\mathrm{d}^2}\frac{\Delta P_\mathrm{e}}{\Delta P_\mathrm{c}},
\end{equation}
Therefore, the constraint of the slope for successful orbital expansion is $a_{\mathrm{cde}} > 1.56$. Next, we analytically solve for $\Delta P_\mathrm{e}/\Delta P_\mathrm{c}$ under the assumption of invariance of the resonance angle and angular momentum conservation.

When three planets with orbital period $P_1, \:P_2$ and $P_3$ are in a $(p,q,r)$ 3BR the period ratio of the three planets satisfies:
\begin{equation}
	\frac{P_\mathrm{3}}{P_\mathrm{2}} = \frac{qP_1}{rP_1-pP_2},
	\label{eq:appA-bce}
\end{equation}
The total angular momentum for the three planets in a 3BR is conserved and can be expressed as:
\begin{equation}
	L_{\mathrm{tot}}=\left(\frac{G^2M^2}{2\pi}\right)^{1/3}\left(m_\mathrm{1}P_\mathrm{1}^{1/3}+m_\mathrm{2}P_\mathrm{2}^{1/3}+m_\mathrm{3}P_\mathrm{3}^{1/3}\right),
	\label{eq:appA-AM}
\end{equation}
Combining angular momentum conservation (\eq{appA-AM}) and the resonance constraint, we can eliminate $P_1$ to obtain:
\begin{equation}
	L_{\mathrm{tot}}=\left(\frac{G^2M^2}{2\pi}\right)^{1/3}\left[m_1\left( \frac{pP_2P_3}{rP_3-qP_2}\right)^{1/3} +m_2 P_2^{1/3} +m_3 P_3^{1/3}\right],
\end{equation}
where the total angular momentum $L_{\mathrm{tot}}$ is conserved. Implicitly differentiating this expression towards $P_2$ gives:
\begin{equation}
	\frac{m_2}{P_2^{2/3}} +\frac{m_3}{P_3^{2/3}} Y_{32} +\frac{p(rP_3^2 -q P_2^2 Y_{32})m_1}{(qP_2-rP_3)^2 \left( \frac{pP_2P_3}{rP_3-qP_2}\right)^{2/3}} = 0,
	\label{eq:Y32}
\end{equation}
where $Y_{32} = \Delta P_3/\Delta P_2$ -- the slope that we are seeking. Here we want to evaluate $\Delta P_\mathrm{3}/\Delta P_\mathrm{2}$ at a certain period ratio $P_\mathrm{3}/P_\mathrm{2}=P_\mathrm{32}$. Solving this equation for $Y_{32}$ we obtain the slope of the expansion:
\begin{equation}
	Y_{32} = \frac{p^{1/3}rP_\mathrm{32}^2m_1 +\left(rP_\mathrm{32}-q\right)^{4/3}P_\mathrm{32}^{2/3}m_2}{p^{1/3}qm_1 -\left(rP_\mathrm{32}-q\right)^{4/3}m_3},
\end{equation}
Similarly, by expressing $P_3$ in \eq{appA-AM} with $P_1$ and $P_2$ and defining $P_{21}=P_2/P_1$, the slope in the period ratio between planets 1 and 2 can be obtained:
\begin{equation}
	Y_{21} = \frac{q^{1/3}pP_\mathrm{21}^2m_3 -\left(r-pP_\mathrm{21}\right)^{4/3}P_\mathrm{21}^{2/3}m_2}{q^{1/3}rm_3+\left(r-pP_\mathrm{21}\right)^{4/3}m_2},
\end{equation}
where $Y_{21} = \Delta P_2/\Delta P_1$.

For the Kepler-221 system, $(p,q,r)=(2,3,5)$, $\Delta P_\mathrm{e}/\Delta P_\mathrm{c} = Y_{32}$, and $P_{32}=3$. Relabelling $(1,2,3)\rightarrow\mathrm{(b,c,e)}$ and substituting $(p,q,r)\rightarrow(2,3,5)$ \eq{Y32} evaluates to:
\begin{equation}
	\frac{\Delta P_\mathrm{e}}{\Delta P_\mathrm{c}}
	= \frac{15m_\mathrm{b} +15.12m_\mathrm{c}}{m_\mathrm{b} -7.27m_\mathrm{e}},
\end{equation}
Therefore we obtain the dependence of $a_{\mathrm{cde}}$ on the mass of planets (also shown in \eq{slope_for}):
\begin{equation}
	a_{\mathrm{cde}} = \frac{4.90m_\mathrm{b}+4.94m_\mathrm{c}}{7.27m_\mathrm{e}-m_\mathrm{b}},
	\label{eq:app-A}
\end{equation}
\Eq{app-A} shows the slope is steeper with more massive planets b and c and shallower with a more massive e. The slope in the simulation is consistent with this analytical expression although marginal differences arise due to the migration of planet d during the orbital expansion phase, which will be further discussed in \App{d-mig}.

\begin{figure}
	\includegraphics[width=\columnwidth]{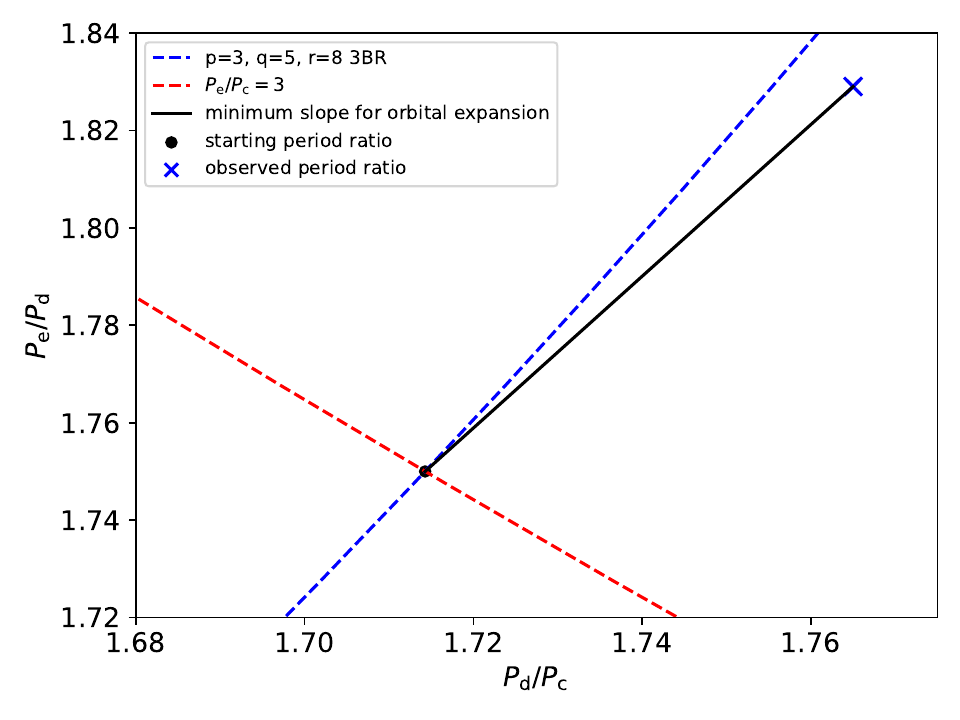}
	\caption{\label{fig:appA} Simplified model of the orbital expansion phase showing the evolution of $P_\mathrm{d}/P_\mathrm{c}$ and $P_\mathrm{e}/P_\mathrm{d}$. The blue dashed line corresponds to the $(3,5,8)$ 3BR for planets c, d, and e. The red dashed line represents the period ratio following $P_\mathrm{e}/P_\mathrm{c}>3$. The blue cross is the observed period ratio and the black line represents the minimum slope of $P_\mathrm{d}/P_\mathrm{c}$ and $P_\mathrm{e}/P_\mathrm{d}$ which ensures successful orbital expansion to the observed period ratio.}
\end{figure}

\section{Origin of the migration of planet d in Phase IV}
\label{app:d-mig}
\begin{figure*}
	\sidecaption
	\includegraphics[width=0.67\textwidth]{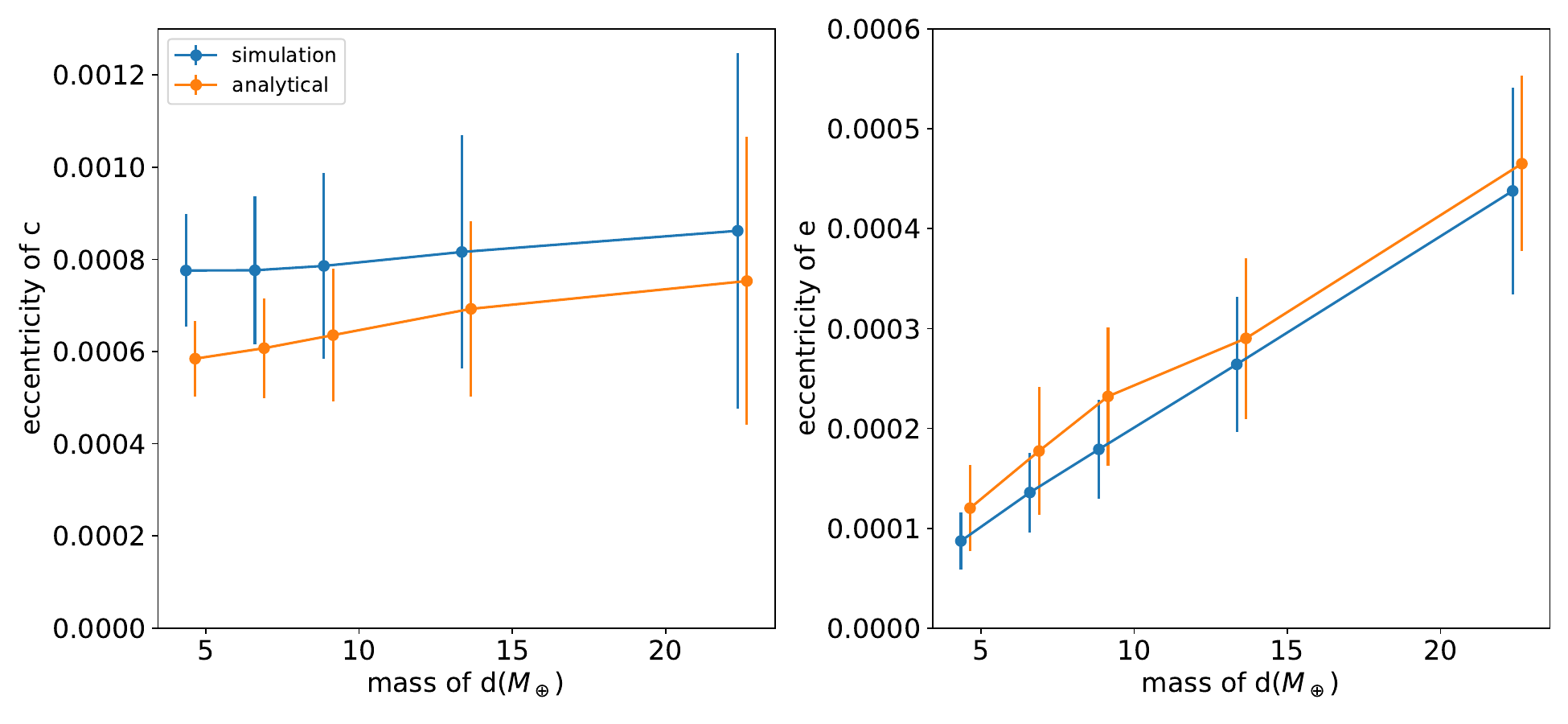}
	\caption{\label{fig:appB_1} Comparison between the eccentricity of planets in the N-body simulation (blue line) and the analytical value due to secular interaction, calculated according to \citet{MurrayEtal1999}. The left panel corresponds to the eccentricity of planet c and the right panel corresponds to planet e. The blue line shows the average and standard deviation of planet eccentricities in the simulation with different masses of d and the yellow line shows the mean and standard deviation of the Laplace-Lagrante theory averaged over random initial conditions sampled at different time. }
\end{figure*}

\begin{figure}
	\includegraphics[width=\columnwidth]{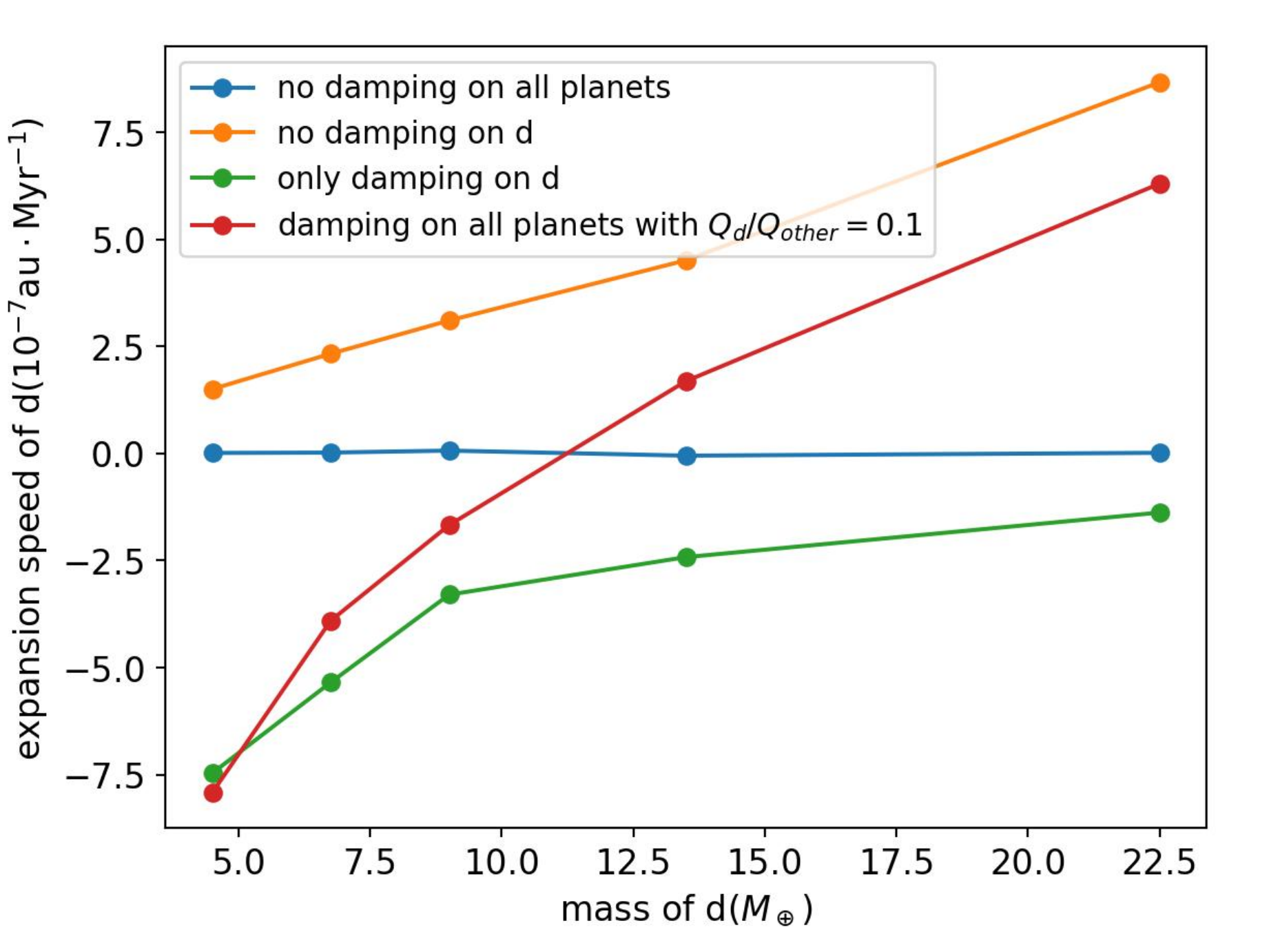}
	\caption{\label{fig:appB_2} Evolution of planet d in the orbital expansion phase with different initial conditions. The damping is applied on planets b, c, and e for the yellow and red line, and on planet d for the green and red line. The tidal $Q$ parameter on different planets follow the relation $Q_\mathrm{d}/Q_\mathrm{other}=0.1$.}
\end{figure}

In panel b of \fg{orb_exp_fail}, the evolution of $P_\mathrm{d}/P_\mathrm{c}$ and $P_\mathrm{e}/P_\mathrm{d}$ in the simulation (colored line) show minor deviation from the analytical evolution (black line) due to the outward migration of planet d. To explain such migration, this section investigates secular interaction between planets \citep{MurrayEtal1999}. Secular interaction contributes to the minimum value of the eccentricity of the planets without resonance interaction. In the orbital expansion phase, tidal dissipation expands the b, c, and e 3BR and conserves angular momentum, while planet d would migrate outward slightly due to angular momentum change. The relative tidal dissipation strength of resonant system expansion depends on the initial eccentricity of planets in resonance (i.e. b, c, and e).

Therefore, we estimate the eccentricity of planets c and e due to their secular interactions with planet d with the Laplace-Lagrange theory \citep{MurrayEtal1999}, in which the evolution of the eccentricity of the planets due to mutual secular interaction is analytically expressed by:
\begin{align}
	 & h_j =e_{j1}\sin\left(g_1t+\beta_1\right)+e_{j2}\sin\left(g_2t+\beta_2\right) \\
	 & k_j =e_{j1}\cos\left(g_1t+\beta_1\right)+e_{j2}\cos\left(g_2t+\beta_2\right) \\
	 & e_j(t) =\left(h_j^2+k_j^2\right)^{1/2}
\end{align}
where $j=1,\;2$. $e_j(t)$ shows the time evolution of the eccentricity of the two planets. Parameters in the formula including $e_{j1},\;e_{j2},\;g_1,\;g_2,\;\beta_1$ and $\beta_2$ depend on the initial mass, semi-major axis, eccentricity, and orbital position of the two planets. We calculate the mean and standard deviation of the eccentricity of planets c and e due to the secular perturbation of planet d averaged over random initial conditions at the start of the orbital expansion phase, and compare it with the mean and average of the eccentricity in the simulation. The result is shown in \fg{appB_1}, in which the blue line represents the simulation result and the orange line represents the analytical result. From the figure, it is clear that the two curves have similar values, indicating that the secular perturbation of planet d contributes most of the eccentricity of planets c and e. The difference between the simulation and analytical value is an outcome of secular and resonance perturbation from other planets. Also, the eccentricity for planets c and e is larger with a more massive planet d.

The dependence of the eccentricity of planets b, c, and e on the mass of d helps explain the evolution of planet d in the post-collision phase, as shown in \fg{appB_2}, which shows the outward or inward migration speed of planet d with different initial conditions.

For the blue line in \fg{appB_2}, planet d migrates neither outward nor inward because no tidal damping is applied on planet d. For the yellow line, tidal damping is added on planets b, c, and e but not on planet d, and planet d would migrate outward at a faster speed with a more massive planet d. This is because the tidal damping on planets b, c, and e would expand the 3BR, increasing the relative period ratio. During this process, tidal dissipation conserves angular momentum, planets b and c would migrate inward and planets d and e would migrate outward slightly due to the angular momentum exchange. A more massive d planet will increase the eccentricity of the other planets, increasing their relative tidal dissipation and the outward migration speed of planet d. Such effect of outward migration explains the difference between the black line and the colored line in panel b of \fg{orb_exp_fail}, in which the simulation (colored line) deviates from the analytical result (black line) because planet d also moves outward during the orbital expansion phase.

Tidal damping is only applied on planet d for the green line in \fg{appB_2}, thus planet d migrates inward. This is because the eccentricity of planet d is already at the minimum value allowed by the secular interaction from other planets, so the tidal damping on planet d would not change its eccentricity but still migrate the planet inward. The tidal dissipation strength is independent of the mass of planet d, and less massive planets are affected more by the dissipation, leading to the mass dependence that the speed of inward migration is faster for the less massive planet d. 

The red line in \fg{appB_2} corresponds to the case where tidal damping is added on all the planets. The value of the red line is similar to the addition of the yellow and blue line combining the two effects, with the less massive planet d migrating inward and the more massive planet d migrating outward.


\end{document}